\documentclass[12pt]{JHEP3}
\usepackage{latexsym}
\usepackage{axodraw}
\usepackage{epsfig}
\usepackage{amsfonts}
\usepackage{array,longtable}

\title{\bf Renormalizable $1/N_f$ Expansion for Field Theories in Extra Dimensions}
\author{D.I.Kazakov$^{1,2}$ and G.S.Vartanov$^{1,3}$\\
$^1$Bogoliubov Laboratory of Theoretical Physics, Joint
Institute for Nuclear Research, Dubna, Russia \\
$^2$Institute for Theoretical and Experimental Physics, Moscow, Russia\\
$^3$University Center, Joint Institute for Nuclear Research,
Dubna, Russia }

\abstract{ We demonstrate how one can construct renormalizable
perturbative expansion in formally nonrenormalizable higher
dimensional field theories. It is based on $1/N_f$-expansion and
results in a logarithmically divergent perturbation theory in
arbitrary high space-time dimension. First, we consider a simple
example of $N$-component  scalar filed theory and then extend this
approach to Abelian and non-Abelian gauge theories with $N_f$
fermions. In the latter case, due to self-interaction of
non-Abelian fields the proposed recipe requires some modification
which, however, does not change the main results. The resulting
effective coupling is dimensionless and is running in accordance
with the usual RG equations. The corresponding beta function is
calculated in the leading order and is nonpolynomial in effective
coupling. It exhibits either UV asymptotically free or IR free
behaviour depending on the dimension of space-time. The original
dimensionful coupling plays a role of a mass and is also
logarithmically renormalized. We analyze also the analytical
properties of a resulting theory and demonstrate that in general
it acquires several ghost states with negative and/or complex
masses. In the former case, the ghost state can be removed by a
proper choice of the coupling. As for the states with complex
conjugated masses, their contribution to physical amplitudes
cancels so that the theory appears to be unitary.}

\begin{document}
\section{Introduction}

Nowadays it is popular to consider theories in extra dimensions as
possible candidates for models of physics beyond the Standard
Model~\cite{Original,review}. Within the braneworld scenario one
assumes that the matter fields are localized at the brane while
the force carriers can travel in the bulk~\cite{inter}. Sometimes
other fields might also live in extra dimensions. This means that
one has higher dimensional QFT at least at short distances.
However, it can hardly be considered as a consistent quantum
theory beyond the tree level because of a lack of renormalizable
perturbative expansion. Indeed, the usual coupling has a negative
dimension, thus leading to power increasing divergencies which are
out of control. In our previous work~\cite{our}, we studied the UV
divergencies in scalar theories in extra dimensions within the
perturbative expansion and demonstrated that although the leading
divergences are governed by the one-loop diagrams even in the
nonrenormalizable case, as was argued in~\cite{Kazakov}, this does
not help to conquer them.

Popular reasoning when dealing with extra dimensional theories
relies on higher energy (string) theory which is supposed to cure
all the UV problems while the low energy one is treated as an
effective theory basically at the tree level. One way to do it is
the Kaluza-Klein approach~\cite{KK}. In this case, one takes the
Fourier transform over the extra dimensions and obtains an
infinite tower of states with quantized masses. Then one has to
sum over all these states. This sum is usually divergent and a
special prescription is needed to regularize it. Doubtfully,
however, that this approach solves the problem of
nonrenormalizability in extra dimensional theories. As was shown
in~\cite{esp}, the properly renormalized four-dimensional theory
never forgets its higher dimensional origin. It has an explicit
cut-off dependence and can only be treated as an effective
theory~\cite{ratt}.

Here we make an attempt to construct renormalizable expansion in such formally
nonrenormalizable theories using the well known technique of the $1/N$
expansion~\cite{Parisi, 1N}, where in the scalar case $N$ is the number of the scalar
field components and in gauge theories it is the number of fermion flavours $N_f$. The
number of colours $N_c$ is kept fixed. This approach was successfully applied to
non-linear sigma-models in 3 dimensions~\cite{Arefeva} and to quantum gravity in four
dimensions~\cite{Tomboulis} both of which are nonrenormalizable by power counting.
Effectively, as we will show below, it leads to higher derivative theories and causes the
usual problems of unitarity, locality and causality. However, these problems could be
overcome though the analysis was performed mainly in the leading order~\cite{Arefeva,
Tomboulis}.

We follow the approach of~\cite{Arefeva, Tomboulis} and apply it
to theories in extra dimensions with the aim to construct
renormalizable and unitary $1/N$ expansion suitable for
perturbative calculations. We first consider scalar higher
dimensional theories as an example~\cite{KVI} and then treat the
gauge theories with fermions in the same way~\cite{KVII}. The
resulting perturbation theory is shown to be renormalizable,
logarithmically divergent in any dimension $D$ and obtains an
effective dimensionless expansion parameter. It is nonpolynomial
in effective coupling, but polynomial in $1/N$ and obeys the usual
properties of renormalizable theory. It might be either UV
asymptotically free or IR free depending on the space-time
dimension $D$. The original dimensionful coupling does not serve
as an expansion parameter anymore and plays the role of mass which
is also logarithmically divergent and multiplicatively
renormalized.

Within the dimensional regularization technique~\cite{reg} we performed the
renormalization procedure in scalar and gauge theories in arbitrary odd space-time
dimension and calculated a few terms of the 1/N expansion. Even dimensions, in principle,
can also be treated by this method; however, they lead to some complications due to the
appearance of log terms.

It is well known that the main problem of the $1/N$ expansion is
to prove unitarity of a resulting theory since the analytical
properties of the effective propagator change. When summing up the
vacuum polarization diagrams to the denominator of a singlet field
one gets an imaginary part  and, in general, additional poles in
the complex plane. These poles correspond to  ghost states with
the wrong metric and negative or complex masses. It is a common
problem in  any realization of the 1/N expansion~\cite{Tomboulis,
Kr, Schnitzer}. Effectively, it leads to higher derivative terms
which may result in dynamical instability~\cite{Nesterenko}. This
is another issue that we do not discuss here. Note, however, the
existence of "benign" quantum mechanical higher-derivative
systems, where the classical vacuum is stable with respect to
small perturbations and the problems appear only at the
nonperturbative level~\cite{Smilga}.  The question of unitarity in
higher derivative gravity in four dimensions was discussed
in~\cite{Tomboulis}, where the role of ghost states was
emphasized. It was also shown~\cite{Antoniadis} that the higher
derivative operators do not always  improve the UV behaviour due
to subtleties in analytical continuation from Minkowski to
Euclidean metric.

Below we consider the unitarity problem in detail and suggest a possible solution which
seems to lead to a unitary theory in physical subspace.

\section{$1/N$ expansion. Scalar theory}

To illustrate the method, we start with the scalar theory. Let us
take the usual $N$ component scalar field theory in $D$
dimensions, where $D$ takes an arbitrary value ($>4$), with the
$\phi^4$ self-interaction. The Lagrangian looks like
\begin{equation}\label{l0}
  {\cal L} = \frac 12 (\partial_\mu \vec{\phi})^2-\frac12 m^2 \vec{\phi}^2 -
  \frac{\lambda}{8N}(\vec{\phi}^2)^2,
\end{equation}
where $N$ is the number of components of $\phi$. We put $N$ into the normalization of the
coupling so that $\lambda$ is fixed while $N\to\infty$. The theory is nonrenormalizable
by power counting, the coupling $\lambda$ has negative dimension $[\lambda]=2-D/2$. It is
useful to rewrite eq.(\ref{l0}) introducing an auxiliary field $\sigma$~\cite{Arefeva}
\begin{equation}\label{s}
 {\cal L} = \frac 12 (\partial_\mu \vec{\phi})^2-\frac12 m^2 \vec{\phi}^2 -
 \frac{1}{2\sqrt{N}}\sigma(\vec{\phi}^2)+\frac{1}{2\lambda} \sigma^2.
\end{equation}
Now one has two fields, one $N$ component and one singlet with triple interaction. Let us
look at the propagator of the $\sigma$ field. At  the tree level it is just "$i\lambda$",
but then one has to take into account the corrections due to the loops of $\phi$ (see
Fig.1).
\begin{center}
\begin{picture}(300,100)(50,-70)
\SetWidth{1.5} \DashLine(0,0)(45,0){3} \SetWidth{0.5} \Text(52,0)[]{=}
\DashLine(60,0)(90,0){3} \Text(98,0)[]{+} \DashLine(105,0)(135,0){3}
\Oval(150,0)(15,15)(360) \DashLine(165,0)(195,0){3} \Text(203,0)[]{+}
\DashLine(210,0)(240,0){3} \Oval(255,0)(15,15)(360) \DashLine(270,0)(300,0){3}
\Oval(315,0)(15,15)(360) \DashLine(330,0)(360,0){3} \Text(373,0)[]{+...}
 \Text(186,-30)[]{Figure 1: The chain of diagrams giving a contribution to the
 $\sigma$ field propagator
 in}
 \Text(63,-48)[]{the zeroth order of $1/N$ expansion}
 \label{prop1}
\end{picture}
\end{center}

If one follows the $N$ dependence of the corresponding graphs, one finds out that it
cancels: they are all of the zeroth order in $1/N$. Thus, one can sum them up and get
\begin{equation}\label{propscalar1}
  \textbf{- - - -} = \textmd{- - - -} (\frac{1}{1- O \textmd{- - -}})
  =\frac{i}{1/\lambda -\Pi(p^2)},
\end{equation}
where the polarization operator $\Pi(p^2)$ depends on $D$. In the
massless case it looks like
\begin{equation}\label{pol}
\Pi(p^2)=-f(D)(-p^2)^{D/2-2}, \ \ \
f(D)=\frac{\Gamma^2(D/2-1)\Gamma(2-D/2)}{2^{D+1}\Gamma(D-2)\pi^{D/2}}.
\end{equation}

The obtained propagator (\ref{propscalar1}) has a typical for
$1/N$ expansion behaviour. Namely, it has a cut starting from
$p^2=0$ (for $m=0$, otherwise from $4m^2$) and poles at negative
or complex $p^2$ depending on $D$. Notice that $f(D)$ is finite
for any odd $D$, despite naive power counting, and diverges for
even $D$. This is due to the use of dimensional regularization:
the one-loop diagrams in odd dimensions are finite since the gamma
function has poles only at integer negative arguments and not at
half-integer ones. This phenomenon can also be understood in using
other regularization techniques. In general one has the UV
divergence which has to be subtracted. This subtraction requires
the redefinition of simple loop diagrams in $D$ dimensions.
However, the number of these diagrams is limited by $[D/2]-1$,
i.e., in 4 and 5 dimensions one has to define 1 diagram, in 6 and
7 dimensions - 2 diagrams, etc. In what follows we assume that
this definition is made a'la dimensional regularization. Moreover,
for simplicity of integration we limit ourselves to odd
dimensions, which allows us to avoid the appearance of the log
terms.

A special issue is the existence of poles in the propagator.
Usually they signal of the appearance of new asymptotic states
which raise the problem of unitarity of a resulting theory. We
address this problem in more detail in Sec.7. Here we just mention
that one can avoid poles on the real axis and have only complex
conjugated pairs. This is enough for integration in Feynman
diagrams.

Thus, we have now the modified Feynman rules: the $\phi$
propagator is the usual one while the $\sigma$ propagator is given
by eq.(\ref{propscalar1}).  One can then construct the diagrams
using these propagators and the triple vertex having in mind that
any closed cycle of $\phi$ gives an additional factor of $N$ and
any vertex gives $1/\sqrt{N}$.

Let us first analyse the degree of divergence. Let us start with the $\phi$ propagator.
If the diagram with two external $\phi$ lines contains $L$ loops, then it has $2L$
vertices, $2L-1$ $\phi$ lines and $L$ $\sigma$ lines. Since each $\sigma$ line now
behaves like $1/p^{D-4}$, the degree of divergence is
\begin{equation}\label{w}
  \omega(G)=LD - (2L-1)2-L(D-4)= 2 !
\end{equation}
for any D. Since this is a propagator, the divergence is proportional to $p^2$ and thus
is reduced to the logarithmic one.

Let us now take the triple vertex. If it has $L$ loops, then one has $2L+1$ vertices,
$2L$ $\phi$ lines and $L$ $\sigma$ lines. Hence, the degree of divergence is
\begin{equation}\label{w2}
  \omega(G)=LD - (2L)2-L(D-4)= 0 !
\end{equation}
for any $D$. Thus, we again have only logarithmic divergence.

At last, consider the $\sigma$ propagator. In $L$ loops it has $2L$ vertices, $2L$ $\phi$
lines and $L-1$ $\sigma$ lines. The degree of divergence is
\begin{equation}\label{w3}
  \omega(G)=LD - (2L)2-(L-1)(D-4)= D-4.
\end{equation}
This means that in odd $D$ it has no global divergence (again we explore the properties
of dimensional regularization) and the only possible divergencies are those of the
subgraphs eliminated by renormalization of $\phi$ and the coupling. To see this, consider
a genuine diagram for the $\sigma$-field propagator which is shown in Fig.2, where the
blobs denote the 1PI vertex or propagator subgraphs.
\begin{center}
\begin{picture}(100,110)(0,-60)
\SetWidth{1.5} \DashLine(0,0)(15,0){3} \DashLine(80,0)(95,0){3} \SetWidth{0.5}
\Vertex(20,0){5} \Vertex(50,25){5} \Vertex(50,-25){5}
\Oval(50,0)(25,30)(0)
\Text(53,-47)[]{Figure 2: General type of the $\sigma$-field propagator}
\end{picture}
\end{center}
After the $R'$ operation\footnote{The $R'$ operation means that we subtract from the
diagram all divergent subgraphs} we do not have any poles in the integrand for the
remaining one-loop integral. What is left is the finite part containing logarithms of
momenta.
 This final integration has the following form:
$$\int\frac{\ln^n(k^2/\mu^2)\ln^m(k^2/p^2)\ln^k(k^2/(k-p)^2)}{k^2(k-p)^2}d^Dk,$$
where $n,m,k$ are some numbers. We ignore here all the masses since they do not
contribute to the UV behaviour. Due to the naive power counting of divergences in
dimensional regularization we obtain the result proportional to $\Gamma(2-D/2)$ which is
finite for any odd $D$. The logarithms can not change this property.

To demonstrate how this works explicitly, we consider a particular example of the
two-loop diagram. The result of the $R'$-operation is shown in Fig.3.
\begin{center}
\begin{picture}(300,120)(55,-70)

\Text(-10,1)[]{$R'$} \SetWidth{1.5} \DashLine(0,0)(20,0){3} \DashLine(80,0)(100,0){3}
\DashCArc(50,30)(25,-150,-30){3} \SetWidth{0.5} \Oval(50,0)(30,30)(0) \Text(109,-1)[]{=}
\SetWidth{1.5} \DashLine(120,0)(140,0){3} \DashLine(200,0)(220,0){3}
\DashCArc(170,30)(25,-150,-30){3} \SetWidth{0.5} \Oval(170,0)(30,30)(0)
\Text(228,-1)[]{--} \Line(240,0)(295,0) \DashCArc(268,0)(30,0,360){3} \SetWidth{1.5}
\DashCArc(267,0)(20,0,180){3} \DashLine(310,0)(330,0){3} \DashLine(390,0)(410,0){3}
\SetWidth{0.5} \Vertex(360,20){3} \Oval(360,0)(20,30)(0)

\Text(204,-55)[]{Figure 3: Demonstration of the global divergence cancellation in the
two-loop diagram}
\end{picture}
\end{center}
After subtracting the divergence in a subgraph we have prior to the last integration
$$\int\frac{d^{D-2\varepsilon}k}{k^2(p-k)^2}[ \Gamma(-1+\varepsilon)
\frac{\Gamma(D/2-1-\varepsilon)
\Gamma(2-\varepsilon)}{\Gamma(D/2-2)\Gamma(D/2+1-2\varepsilon)}\frac{1}{(k^2)^\varepsilon}
+\frac{1}{\varepsilon}\frac{\Gamma(D/2-1)}{\Gamma(D/2-2)\Gamma(D/2+1)} ]\ .$$
The pole
terms in the integrand cancel and expanding it over $\varepsilon$ one gets $\log(k^2)$.
It is, however, easier to integrate it without expanding over $\varepsilon$ which gives
$$\Gamma(-1+\varepsilon)\frac{\Gamma^2(D/2-1-\varepsilon)\Gamma(2-\varepsilon)
\Gamma(D/2-1-2\varepsilon)\Gamma(2+2\varepsilon-D/2)
}{\Gamma(D/2-2)\Gamma(D/2+1-2\varepsilon)\Gamma(1+\varepsilon)
\Gamma(D-2-3\varepsilon)}\frac{(p^2)^{D/2-2}}{(p^2)^{2\varepsilon}}$$
$$+\frac{1}{\varepsilon}\ \frac{\Gamma(D/2-1)}{\Gamma(D/2-2)\Gamma(D/2+1)}
\frac{\Gamma^2(D/2-1-\varepsilon)\Gamma(2+\varepsilon-D/2)}{\Gamma(D-2-2\varepsilon)}
\frac{(p^2)^{D/2-2}}{(p^2)^{\varepsilon}} \ = \ O(1).$$ Thus, after the $R'$ operation
the diagram is finite and we do not need the $\sigma$ field renormalization.

This way one gets the perturbative expansion with only logarithmic divergences. This is
not expansion over dimensionful coupling $\lambda$ but rather $1/N$ expansion with
dimensionless parameter.

\section{Properties of the 1/N expansion}

Consider now the leading order calculations. We start with the $1/N$ terms for the
propagator of $\phi$ and the triple vertex. One has the diagrams shown in Fig.4. Notice
that besides the one-loop diagrams in the same order of the $1/N$ expansion one has the
two-loop diagram for the vertex.
\begin{center}
\begin{picture}(250,130)(0,-55)
\Line(0,0)(50,0) \SetWidth{1.5} \DashCArc(25,0)(15,0,180){3} \SetWidth{0.5}
\Text(25,-10)[]{a}

\SetWidth{1.5} \DashLine(110,55)(110,35){3} \SetWidth{0.5} \Line(80,-0)(110,35)
\Line(140,-0)(110,35) \SetWidth{1.5} \DashLine(92,10)(128,10){3} \SetWidth{0.5}
\Text(110,-10)[]{b}

\Line(170,0)(250,0) \SetWidth{1.5} \DashLine(190,0)(190,25){3}
\DashLine(230,0)(230,25){3} \SetWidth{0.5} \Line(230,25)(190,25) \Line(190,25)(210,45)
\Line(210,45)(230,25) \SetWidth{1.5} \DashLine(210,45)(210,60){3} \SetWidth{0.5}
\Text(210,-10)[]{c} \Text(119,-30)[]{Figure 4: The leading order diagrams giving a
contribution to the $\phi$ field propagator} \Text(3,-43)[]{and the triple vertex  in
$1/N$ expansion}
\end{picture}
\end{center}

Let us start with the diagram a). One has
$$
I_a \sim  \int \frac{d^{D'}k}{(2\pi)^DN}\frac{1}{[(k-p)^2-m^2][1/\lambda -\Pi(k^2)]}, \ \
D'=D-2\varepsilon .
$$
Since we are interested in the UV behaviour we can omit the mass from the $\phi$ field
propagator and "$1/\lambda$" from the $\sigma$ field propagator and take the massless
limit of the polarization operator $\Pi(k^2)$ . We will restore them when discussing the
analytical properties. Then the UV asymptotics is given by
$$I_a\  \Rightarrow\  \int \frac{d^{D'}k}{(2\pi)^DNf(D)}\frac{1}{(k-p)^2(-k^2)^{D/2-2}}.$$
One can see that the original coupling $\lambda$ plays the role of inverse mass and drops
out from the UV expression. What is left is a dimensionless $1/N$ term.

Calculating the singular parts of the diagrams of Fig.4 in  dimensional regularization
with $D'=D-2\varepsilon$ one finds
\begin{eqnarray}\label{sing}
Diag.a &\Rightarrow& \frac{1}{\varepsilon N}A,\ \ \ Diag.b\ \Rightarrow\
\frac{1}{\varepsilon N}B,\ \ \
Diag.c \ \Rightarrow\ \frac{1}{\varepsilon N}C,\\
&& \nonumber \\ &&\hspace*{-2.5cm}
A=\frac{2\Gamma(D-2)}{\Gamma(D/2-2)\Gamma(D/2-1)\Gamma(D/2+1)\Gamma(2-D/2)}, \ \ \
B=\frac{D}{4-D}A, \ \ \ C=\frac{D(D-3)}{4-D}A. \nonumber
\end{eqnarray}
The corresponding renormalization constants in the $\overline{MS}$ scheme then are
\begin{eqnarray}\label{z}
Z_2^{-1}&=&1-\frac{1}{\varepsilon}\frac{A}{N},\\
Z_1&=&1-\frac{1}{\varepsilon}\frac{B+C}{N}.
\end{eqnarray}
There is no any coupling in these formulas, its role is played by $1/N$ which is
therefore infinitely renormalized. This seems to be unsatisfactory and to overcome this
problem we introduce a new dimensionless coupling $h$ associated with the triple vertex
(and not with the $\sigma$ propagator) as
$$ {\cal L}_{int} = -\frac{\sqrt{h}}{2\sqrt{N}}\sigma \vec{\phi}^2.$$
Then in the leading order in $1/N$ the renormalization constants and the coupling take
the form
\begin{eqnarray}\label{r1}
Z_2^{-1}&=&1-\frac{h}{\varepsilon}\frac{A}{N},\\
Z_1&=&1-\frac{h}{\varepsilon}\frac{B}{N}-\frac{h^2}{\varepsilon}\frac{C}{N}, \\
h_B &=& (\mu^2)^\varepsilon h
Z_1^2Z_2^{-2}=h\left(1-\frac{h}{\varepsilon}\frac{2(A+B)}{N}-
\frac{h^2}{\varepsilon}\frac{C}{N}\right).
\end{eqnarray}

This is not, however, the final expression. To see this, we consider the next order of
the $1/N$ expansion. The corresponding diagrams for the $\phi$ propagator are shown in
Fig.5. Again one can see that the $1/N^2$ terms contain not only the two-loop diagrams
but also the three- and even four-loop ones.

\begin{center}
\begin{picture}(400,235)(0,-170)

\Line(0,0)(80,0) \SetWidth{1.5} \DashCArc(40,0)(20,0,180){3} \DashCArc(40,0)(30,0,180){3}
\SetWidth{0.5} \Line(100,0)(180,0) \Text(40,-30)[]{a} \SetWidth{1.5}
\DashCArc(130,0)(20,0,180){3} \DashCArc(150,0)(20,180,360){3} \SetWidth{0.5}
\Text(140,-30)[]{b} \Line(200,0)(280,0) \SetWidth{1.5} \DashCArc(220,0)(10,0,180){3}
\DashCArc(260,0)(10,0,180){3} \SetWidth{0.5} \Line(300,0)(380,0) \Text(240,-30)[]{c}
\SetWidth{1.5} \DashLine(320,0)(320,45){3} \DashLine(340,0)(340,30){3}
\DashLine(360,0)(360,45){3} \SetWidth{0.5} \Line(320,45)(360,45) \Line(320,45)(340,30)
\Line(340,30)(360,45) \Text(340,-30)[]{d} \Line(0,-100)(100,-100) \Oval(50,-65)(15,20)(0)
\Text(50,-120)[]{e} \SetWidth{1.5} \DashCArc(50,-100)(40,120,180){3}
\DashCArc(50,-100)(40,0,60){3} \DashCArc(50,-45)(15,-140,-40){3} \SetWidth{0.5}
\Line(120,-100)(220,-100) \Oval(170,-65)(15,20)(0) \SetWidth{1.5}
\DashCArc(170,-100)(40,120,180){3} \DashCArc(170,-100)(40,0,60){3}
\DashLine(170,-50)(170,-80){3} \SetWidth{0.5} \Line(250,-100)(350,-100)
\Text(170,-120)[]{f} \SetWidth{1.5} \DashCArc(280,-100)(20,90,180){3}
\DashCArc(320,-100)(20,0,90){3} \SetWidth{0.5} \Line(280,-80)(290,-70)
\Line(280,-80)(290,-90) \Line(290,-70)(290,-90) \SetWidth{1.5}
\DashLine(290,-70)(310,-70){3} \DashLine(290,-90)(310,-90){3} \SetWidth{0.5}
\Line(310,-70)(320,-80) \Line(310,-90)(320,-80) \Line(310,-70)(310,-90)
\Text(300,-120)[]{g}

\Text(192,-140)[]{Figure 5: The second order diagrams giving a contribution
 to the $\phi$ field propagator}
\Text(22,-152)[]{in $1/N$ expansion}
\end{picture}
\end{center}

All these diagrams are double logarithmically divergent, i.e., contain both single and
double poles in dimensional regularization. We calculate the leading double pole after
subtraction of the divergent subgraphs, i.e. perform  the $R'$-operation. The answer is:
\begin{eqnarray}\label{sing2}
Diag.a &\Rightarrow& -\frac{1}{\varepsilon^2 N^2}\frac 12A^2h^2,\ \ \ Diag.b\
\Rightarrow\ -\frac{1}{\varepsilon^2 N^2}ABh^2,\ \ \
Diag.c \ \Rightarrow\ -\frac{1}{\varepsilon^2 N^2}A^2, \nonumber \\
Diag.d &\Rightarrow& -\frac{1}{\varepsilon^2 N^2}\frac 43 ACh^3,\ \ \ Diag.e\
\Rightarrow\ -\frac{1}{\varepsilon^2 N^2}\frac 23 A^2h^3,\ \ \ Diag.f \ \Rightarrow\
-\frac{1}{\varepsilon^2 N^2}\frac 23 ABh^3,\nonumber \\ Diag.g & \Rightarrow&
-\frac{1}{\varepsilon^2 N^2}ACh^4.
\end{eqnarray}
Here we face a problem, namely, subtracting the divergent subgraphs in the graphs e-g, we
get the diagram which is absent in our expansion, since it is already included in our
bold $\sigma$ line (see Fig.6).
\begin{center}
\begin{picture}(110,95)(0,-35)
\Line(0,0)(100,0) \Oval(50,35)(15,20)(0) \SetWidth{1.5} \DashCArc(50,0)(40,120,180){3}
\DashCArc(50,0)(40,0,60){3} \SetWidth{0.5} \Text(50,-20)[]{Figure 6: The "forbidden" loop
diagram}
\end{picture}
\end{center}
There would be no problem unless this diagram is needed to match the so-called pole
equations~\cite{Hooft} which allow one to calculate the higher order poles in the Z
factors from the single one. However, if we include this diagram in the $\sigma$ line, it
will not change the latter, except for the additional $h$ factor coming from the vertex
and not compensated by the propagator. Apparently, one can continue this insertion
procedure and add any number of such loops not changing the order of $1/N$ expansion. The
result is the sum of a geometrical progression
$$\frac{1}{1+h},$$
which should multiply every $\sigma$ line.  Altogether this leads to the following
effective Lagrangian for UV $1/N$ perturbation theory
\begin{equation}\label{ef}
 {\cal L}_{eff} = \frac 12 (\partial_\mu \vec{\phi})^2-
 \frac{\sqrt{h}}{2\sqrt{N}}\sigma(\vec{\phi}^2)+\frac{1}{2\lambda} \sigma^2
 +\frac 12 f(D)\sigma (\partial^2)^{D/2-2}\sigma (1+h).
\end{equation}

Having all this in mind we come to the final expressions for the Z factors within the
$1/N$ expansion:
\begin{eqnarray}\label{newscalar}
Z_1&=&1-\frac{1}{\varepsilon N}\left(\frac{B h}{1+h}+\frac{Ch^2}{(1+h)^2}\right)
+O(\frac{1}{N^2}), \\
Z_2^{-1}&=&1-\frac{1}{\varepsilon N}\frac{Ah}{1+h}+\frac{1}{\varepsilon^2 N^2}\left(
\frac 32\frac{A^2h^2}{(1+h)^2}+\frac{ABh^2}{(1+h)^2}+\frac
23\frac{A^2h^3}{(1+h)^3}\right.\nonumber \\ &&\left. + \frac 23\frac{ABh^3}{(1+h)^3}
+\frac 43\frac{ACh^3}{(1+h)^3}+\frac{ACh^4}{(1+h)^4}\right)+O(\frac{1}{\varepsilon N^2}).
\label{new2}
\end{eqnarray}

\section{1/N$_f$ expansion. QED}

Let us consider now the usual QED with $N_f$ fermion fields in $D$ dimensions, where $D$
takes an arbitrary odd value. The Lagrangian looks like
\begin{eqnarray}\label{l}
  {\cal L} &=&
    - \frac 14 (\partial_\mu A_{\nu} - \partial_\nu A_{\mu})^2 -
    \frac {1}{2\alpha} (\partial_{\mu}A_{\mu})^2 + i \bar{\psi_i} \hat{\partial}
  \psi_i - m \bar{\psi_i} \psi_i
  + \frac{e}{\sqrt{N_f}} \bar{\psi_i} \hat{A} \psi_i.
\end{eqnarray}

According to the general strategy, we now have to consider the
photon propagator. Since due to the gauge invariance the
polarization operator is transverse, it is useful to consider a
transverse (Landau) gauge. This is not necessary but simplifies
the calculations. Then in the leading order of the 1/N expansion
one has the following sequence of bubbles (see Fig.7)
\begin{center}
\begin{picture}(300,100)(50,-70)
\SetWidth{1.5} \Photon(0,0)(45,0){1}{6} \SetWidth{0.5} \Text(52,0)[]{=}
\Photon(60,0)(90,0){1}{6} \Text(98,0)[]{+} \Photon(105,0)(135,0){1}{6}
\ArrowArc(150,0)(15,0,180) \ArrowArc(150,0)(15,180,360) \Photon(165,0)(195,0){1}{6}
\Text(203,0)[]{+} \Photon(210,0)(240,0){1}{6} \ArrowArc(255,0)(15,0,180)
\ArrowArc(255,0)(15,180,360) \Photon(270,0)(300,0){1}{6} \ArrowArc(315,0)(15,0,180)
\ArrowArc(315,0)(15,180,360) \Photon(330,0)(360,0){1}{6} \Text(373,0)[]{+...}
 \Text(186,-30)[]{Figure 7: The chain of diagrams giving a contribution to the
 $A$ field propagator in}
 \Text(75,-48)[]{the zeroth order of the $1/N_f$ expansion}
 \label{prop2}
\end{picture}
\end{center}
summed up into a geometrical progression. This is nothing more than the renormalon
chain~\cite{renormalon}. The resulting photon propagator takes the form similar to that
for an auxiliary field $\sigma$ in scalar case
\begin{equation}\label{p}
  D_{\mu\nu}(p) = -\frac{i}{p^2}\left(g^{\mu\nu}-\frac{p^\mu p^\nu}{p^2}\right)
  \frac{1}{1 + e^2 f(D)(-p^2)^{D/2-2}},
\end{equation}
where
$$f(D)=\frac{\Gamma^2(D/2)\Gamma(2-D/2)}{2^{D-[D/2]-1}\Gamma(D)\pi^{D/2}}$$
and we put $m=0$ for simplicity.

This practically coincides with the expression obtained in scalar theory and all the
following steps just repeat those in the latter. We change the normalization of the gauge
field $A_\mu\to A_\mu/e$ and  introduce the dimensionless coupling $h$ associated with
the triple vertex, so the effective Lagrangian takes the form
\begin{eqnarray}\label{eff}
\nonumber  {\cal L}_{eff} &=& - \frac 14 F_{\mu\nu}\left( \frac{1}{e^2}
  + f(D)(\partial^2)^{D/2-2} (1+h)
  \right) F_{\mu\nu} - \frac {1}{2\alpha e^2} (\partial_{\mu}A_{\mu})^2 \\
&&   + i \bar{\psi_i} \hat{\partial}  \psi_i - m \bar{\psi_i} \psi_i +
\frac{\sqrt{h}}{\sqrt{N_f}} \bar{\psi_i} \hat{A} \psi_i.
\end{eqnarray}
This new dimensionless coupling $h$ enters into the gauge
transformation and plays the role of a gauge charge. The old
coupling $e$, on the contrary, is dimensionful and acts as a mass
parameter in a gauge propagator. Since the coupling constant $h$
is dimensionless the effective Lagrangian (\ref{eff}) when
omitting the first term is conformal as considered
in~\cite{Anselmi} where the theory was taken in $D=3$.

Again, one has the modified Feynman rules with the photon propagator that decreases in
the Euclidean region like $1/(p^2)^{D/2-1}$, thus improving the UV behaviour in a theory.
The only divergent graphs are those of the fermion propagator and the triple vertex. They
are both logarithmically divergent for any odd D. The photon propagator is genuinely
finite and may contain divergencies only in subgraphs. One basically has the same graphs
as in  a scalar theory but with solid lines being the fermion ones and the dashed lines
being the photon one.

The only difference (or simplification) comes from the Furry theorem and the gauge
invariance. Namely,  all triangles with three photon external lines vanish due to the
Furry theorem and the gauge invariance which connects the fermion propagator with the
triple vertex implies that $Z_1=Z_2$. This relation holds in the $1/N_f$ expansion like
in the usual PT. Thus, using the notation of a previous section, in the leading order one
has
\begin{equation}
A=\frac{\Gamma(D)(D-1)(2-D/2)}{2^{[D/2]+1} \Gamma(2-D/2)
\Gamma(D/2+1) \Gamma^2(D/2)}, \ B=-A, \ C=0.
\end{equation}
The same results were obtained in~\cite{Gracey1} where the author calculated the
anomalous dimensions at the D-dimensional critical point where the fields obey asymptotic
scaling and are conformal. This leads to the following renormalization constants in the
leading order in $1/N_f$:
\begin{eqnarray}\label{r2}
Z_1=Z_2&=&1+\frac{1}{\varepsilon N_f}\frac{Ah}{(1+h)},\ \ Z_3=1
\end{eqnarray}
and, consequently, $h_B = h$. Hence, in odd-dimensional QED in the leading order of the
$1/N_f$ expansion one does not need the coupling constant renormalization; only the wave
function renormalization remains. This means that the coupling is not running.

In the second order one again has the same diagrams as in a scalar theory but with
vanishing triangles. The renormalization constant in the second order is also essentially
simplified compared to the scalar case and looks like
\begin{equation}\label{new}
Z_1=1+\frac{1}{\varepsilon N_f}\frac{A h}{1+h} +\frac{1}{\varepsilon^2 N_f^2} \frac
12\frac{A^2h^2}{(1+h)^2}+O(\frac{1}{\varepsilon N_f^2}).
\end{equation}

Like in the scalar case the original dimensionful coupling $e$ is
not an expansion parameter anymore, but plays a role of a mass and
is multiplicatively logarithmically renormalized. The leading
order diagrams  are shown in Fig.8.
\begin{center}
\begin{picture}(400,120)(0,-160)

\Oval(50,-65)(15,20)(0) \ArrowArc(50,-65)(15,89,91) \ArrowArc(50,-65)(15,269,271)
\Text(50,-90)[]{a} \SetWidth{1.5} \Photon(10,-65)(30,-65){1}{3}
\Photon(70,-65)(90,-65){1}{3} \PhotonArc(50,-45)(15,-143,-37){1}{5} \SetWidth{0.5}
\Line(46,-56)(54,-64) \Line(54,-56)(46,-64)

\Oval(170,-65)(15,20)(0) \ArrowArc(170,-65)(16.5,44,46) \ArrowArc(170,-65)(16.5,134,136)
\ArrowArc(170,-65)(16.5,224,226) \ArrowArc(170,-65)(16.5,314,316) \SetWidth{1.5}
\Photon(130,-65)(150,-65){1}{3} \Photon(190,-65)(210,-65){1}{3}
\Photon(170,-50)(170,-80){1}{3} \SetWidth{0.5} \Line(166,-61)(174,-69)
\Line(174,-61)(166,-69) \Text(170,-90)[]{b}

\SetWidth{1.5} \Photon(260,-65)(280,-65){1}{3} \Photon(320,-65)(340,-65){1}{3}
\SetWidth{0.5} \ArrowLine(280,-65)(290,-55) \ArrowLine(290,-75)(280,-65)
\ArrowLine(290,-55)(290,-75) \SetWidth{1.5} \Photon(290,-55)(310,-55){1}{3}
\Photon(290,-75)(310,-75){1}{3} \SetWidth{0.5} \Line(296,-51)(304,-59)
\Line(304,-51)(296,-59) \ArrowLine(320,-65)(310,-55) \ArrowLine(310,-75)(320,-65)
\ArrowLine(310,-55)(310,-75) \Text(300,-90)[]{c}

\Text(192,-110)[]{Figure 8: The first order diagrams giving a
contribution
 to the $1/e^2$ renormalization}
\Text(34,-122)[]{in the $1/N_f$ expansion}
\end{picture}
\end{center}\vspace{-1cm}
They give the following contribution:
\begin{eqnarray}\label{lambda}
Diag.a &\Rightarrow& \frac{h^2}{\varepsilon N_f (1+h)^2}F,\ \ Diag.b\ \Rightarrow\
\frac{h^2}{\varepsilon N_f (1+h)^2}E,\ \
Diag.c \ \Rightarrow\ 0,  \\
&& \nonumber \\ &&\hspace*{-2.5cm} F=\frac{\Gamma(D+1)(D/2-1)(D-1)^2(2-D/2)}{2^{D/2+1}
\Gamma(2-D/2) \Gamma(D/2+2) \Gamma^2(D/2)}, \ E=-\frac{D^2+D/2-9}{D/2(D/2-1)(D-1)}F.
\nonumber
\end{eqnarray}
So one has
\begin{equation}
Z_{1/ e^2} \ = \ 1 - \frac{1}{\varepsilon N_f} \left( \frac{(F+E)h^2}{(1+h)^2}
\right)+O(\frac{1}{N_f^2}).
\end{equation}

\section{1/N$_f$ expansion. QCD}


Consider now a non-Abelian theory with $N_f$ fermions. Notice that in QCD, contrary to
QED, all Feynman diagrams contain group factors so that the actual expansion parameter
becomes $N_c/N_f$, thus requiring that this ratio is small. At the same time, to preserve
asymptotic freedom in 4 dimensions one needs $N_c/N_f>2/11$. So one has some interval
where the $N_c/N_f$ expansion might be valid. Of course, in non-Abelian theories the
$1/N_c$ expansion would be preferable, since it accumulates the interactions of the gauge
fields, however, in this case already the lowest approximation consists of all planar
diagrams and is not known~\cite{Hooft1}.

In the non-Abelian case one has some novel features due to the presence of the triple and
quartic gauge vertices and the ghost fields. Similar to (\ref{l}) we write down the
Lagrangian for the gauge fields and $N_f$ fermions as
\begin{eqnarray}\label{ll}
  {\cal L} &=& - \frac 14 (F_{\mu \nu}^a)^2 - \frac {1}{2\alpha} (\partial_{\mu}A_{\mu}^a)^2 + i \bar{\psi_i} \hat{\partial}
  \psi_i - m \bar{\psi_i} \psi_i
  + \frac{g}{\sqrt{N_f}} \bar{\psi_i} \hat{A^a}T^a \psi_i + \partial_{\mu} \bar{c^a}
  {D}_{\mu} c_a,
  \nonumber
\end{eqnarray}
where
$$F_{\mu\nu}^a \ = \ \partial_\mu A_{\nu}^a -
\partial_\nu A_{\mu}^a + \frac{g}{\sqrt{N_f}}f^{abc}A_{\mu}^bA_{\nu}^c, \ \ \ D_\mu = \partial_\mu
+\frac{g}{\sqrt{N_f}}[A_\mu,\ ]$$
 Like in QED we choose the Landau gauge and sum up the fermion bubble diagrams into the denominator of the
gauge field propagator
\begin{equation}\label{ppp}
G^{ab}_{\mu\nu} \ = \ -\frac{i \delta^{ab}}{p^2} \frac{(g^{\mu\nu} -
\frac{p^{\mu}p^{\nu}}{p^2})}{1 + g^2 f(D)(-p^2)^{D/2-2}},
\end{equation}
where the coefficient $f(D)$ differs from the Abelian case only by the colour factor
$T(R)$
$$f(D)=\frac{\Gamma^2(D/2)\Gamma(2-D/2)}{2^{D-[D/2]-1}\Gamma(D)\pi^{D/2}}T(R)$$
and again we put $m=0$ for simplicity.

In the non-Abelian case, contrary to the Abelian one, one has the
triple and quartic self-interaction of the gauge fields. These
vertices, which are suppressed by $1/\sqrt{N_f}$ and $1/N_f$,
respectively, obtain loop corrections of the same order in
$1/N_f$. The effective vertices in the leading order are given by
the diagrams shown in Fig. 9 and 10.\vspace{-1cm}
\begin{center}
\begin{picture}(150,100)(50,-40)
\Gluon(0,0)(0,30){1}{6} \Gluon(0,0)(20,-30){1}{9} \Gluon(0,0)(-20,-30){1}{9}
\Vertex(0,0){5} \Text(25,0)[]{=} \Gluon(80,0)(80,30){1}{6} \Gluon(80,0)(50,-30){1}{9}
\Gluon(80,0)(110,-30){1}{9} \Text(115,0)[]{+} \ArrowLine(155,10)(170,-10)
\ArrowLine(170,-10)(140,-10) \ArrowLine(140,-10)(155,10) \Gluon(155,10)(155,30){1}{6}
\Gluon(170,-10)(180,-30){1}{6} \Gluon(140,-10)(130,-30){1}{6}
 \Text(107,-50)[]{Figure 9: The diagrams giving a contribution to the
 $A^3$ term in the zeroth order}
 \Text(-43,-68)[]{of the $1/N_f$ expansion}
 \label{prop3}
\end{picture}
\end{center}\vspace{2.7cm}

\begin{center}
\begin{picture}(150,60)(50,-70)

\Gluon(0,0)(30,-30){1}{9} \Gluon(0,0)(30,30){1}{9} \Gluon(0,0)(-30,30){1}{9}
\Gluon(0,0)(-30,-30){1}{9} \Vertex(0,0){5} \Text(35,0)[]{=} \Gluon(80,0)(50,-30){1}{9}
\Gluon(80,0)(50,30){1}{9} \Gluon(80,0)(110,30){1}{9} \Gluon(80,0)(110,-30){1}{9}
\Text(115,0)[]{+} \ArrowLine(150,15)(180,15) \ArrowLine(180,15)(180,-15)
\ArrowLine(180,-15)(150,-15) \ArrowLine(150,-15)(150,15) \Gluon(150,15)(125,30){1}{6}
\Gluon(180,15)(205,30){1}{6} \Gluon(150,-15)(125,-30){1}{6}
\Gluon(180,-15)(205,-30){1}{6}

\Text(107,-50)[]{Figure 10: The diagrams giving a contribution to
the $A^4$ term in the zeroth order}
 \Text(-43,-68)[]{of the $1/N_f$ expansion}
 \label{prop4}
\end{picture}
\end{center}

Thus, besides the modification of the gauge propagator one has the modified vertices. The
effective Lagrangian in the case of vertices is not given by a simple local expression
due to complexity of the loop diagrams. So we keep it in the form of the diagrams which
have to be evaluated in integer dimension. Due to the rules of dimensional regularization
they are finite for any odd $D$, otherwise one has to redefine them. What is crucial,
however, is that there are only three diagrams which have to be redefined. Hence, after
rescaling the gauge field $A_\mu \to A_\mu/g$ one obtains the following effective
Lagrangian: \vspace{-1cm}
\begin{eqnarray}\label{lll1}
  {\cal L}_{eff} &=& - \frac{1}{4g^2} (F_{\mu \nu}^a)^2
\begin{picture}(100,50)(-30,-5)\Text(-7,0)[]{-- (} \Gluon(0,0)(10,0){1}{3}
\Oval(20,0)(10,10)(0) \Gluon(30,0)(40,0){1}{3} \Text(55,0)[]{+} \Oval(80,0)(10,10)(0)
\Gluon(60,-10)(70,0){1}{3}\Gluon(100,-10)(90,0){1}{3}\Gluon(80,10)(80,20){1}{3}
\Text(110,0)[]{+} \Oval(140,0)(10,10)(0)
\Gluon(120,-10)(130,-5){1}{3}\Gluon(120,10)(130,5){1}{3}\Gluon(160,-10)(150,-5){1}{3}
\Gluon(160,10)(150,5){1}{3}\Text(170,0)[]{)}
\end{picture}\\
  &-&\frac {1}{2\alpha g^2} (\partial_{\mu}A_{\mu}^a)^2 + i \bar{\psi_i} \hat{\partial}
  \psi_i - m \bar{\psi_i} \psi_i
  + \frac{1}{\sqrt{N_f}} \bar{\psi_i} \hat{A^a}T^a \psi_i + \partial_{\mu} \bar{c^a}
  {D}_{\mu} c_a,
  \nonumber
\end{eqnarray}
Notice that dimensionful coupling $g$ drops from all terms except for the first one and
is not an expansion parameter anymore.

Calculating the degree of divergence after summing up the diagrams of the zeroth order,
similar to the scalar case and QED, one has only four types of logarithmically divergent
diagrams: the fermion and the ghost propagators, the fermion-gauge-vertex and
ghost-gauge-ghost vertex. The gauge propagator as well as pure gauge vertices are finite
and may contain only divergent subgraphs.

The next step is the introduction of a dimensionless coupling $h$. Here one should be
accurate since this coupling enters not only into the triple gauge-fermion vertex, but
due to the gauge invariance should be present in gauge and gauge-ghost vertices. It
should be the same in all three of them. In the case of a gauge theory, the coupling $h$
enters the gauge transformation and acts as a gauge charge of the fermion and gauge
fields.

When constructing the Feynman diagrams, one reproduces the
one-loop cycles that are already present in the effective
Lagrangian (\ref{lll1}) but with additional factors h. In the
scalar or QED case, this happened only  for the propagator, but
here it is also true for the vertices. As a result, the final
expression for the effective Lagrangian takes the form
\vspace{-1cm}
\begin{eqnarray}\label{lll}
  {\cal L}_{eff} &=& - \frac{1}{4g^2} (F_{\mu \nu}^a)^2
\begin{picture}(100,50)(-30,-5)\Text(-7,0)[]{-- (} \Gluon(0,0)(10,0){1}{3}
\Oval(20,0)(10,10)(0) \Gluon(30,0)(40,0){1}{3} \Text(55,0)[]{+} \Oval(80,0)(10,10)(0)
\Gluon(60,-10)(70,0){1}{3}\Gluon(100,-10)(90,0){1}{3}\Gluon(80,10)(80,20){1}{3}
\Text(110,0)[]{+} \Oval(140,0)(10,10)(0)
\Gluon(120,-10)(130,-5){1}{3}\Gluon(120,10)(130,5){1}{3}\Gluon(160,-10)(150,-5){1}{3}
\Gluon(160,10)(150,5){1}{3}\Text(185,0)[]{)\ (1+h)}
\end{picture}\\
  &-&\frac {1}{2\alpha g^2} (\partial_{\mu}A_{\mu}^a)^2 + i \bar{\psi_i} \hat{\partial}
  \psi_i - m \bar{\psi_i} \psi_i
  + \frac{\sqrt{h}}{\sqrt{N_f}} \bar{\psi_i} \hat{A^a}T^a \psi_i + \partial_{\mu} \bar{c^a}
  {D}_{\mu} c_a,
  \nonumber
\end{eqnarray}
where
$$F_{\mu\nu}^a \ = \ \partial_\mu A_{\nu}^a - \partial_\nu A_{\mu}^a + \frac{\sqrt{h}}{\sqrt{N_f}}f^{abc}A_{\mu}^bA_{\nu}^c \
\ ,\ \ D_{\mu}c_a \ = \
\partial_{\mu}c_a+\frac{\sqrt{h}}{\sqrt{N_f}}f^{abc}A^b_{\mu}c^c.$$

Consider now the leading order calculations. We start with the
$1/N_f$ terms for the fermion and the triple fermion-gauge-fermion
vertex. The diagrams are shown in Fig.11. The first two are the
same as in QED. The third diagram contains new effective vertex
which includes the usual triple vertex and the fermion triangle.
The usual vertex does not give a contribution since it is finite
by a simple power counting. At the same time, the fermion triangle
is momentum dependent and the resulting diagram is logarithmically
divergent.
\begin{center}
\begin{picture}(250,130)(0,-55)
\ArrowLine(-10,0)(10,0) \ArrowLine(10,0)(40,0) \ArrowLine(40,0)(60,0) \SetWidth{1.5}
\GlueArc(25,0)(15,0,180){1}{6} \SetWidth{0.5} \Text(25,-10)[]{a}

\SetWidth{1.5} \Gluon(110,55)(110,35){1}{3} \SetWidth{0.5} \ArrowLine(80,-0)(110,35)
\ArrowLine(110,35)(140,-0) \SetWidth{1.5} \Gluon(89,10)(131,10){1}{6} \SetWidth{0.5}
\Text(110,-10)[]{b}

\ArrowLine(170,0)(190,0) \ArrowLine(190,0)(230,0) \ArrowLine(230,0)(250,0) \SetWidth{1.5}
\Gluon(190,0)(210,35){1}{6} \Gluon(230,0)(210,35){1}{6}
\Vertex(210,35){3} \Gluon(210,35)(210,60){1}{3} \SetWidth{0.5}
\Text(210,-10)[]{c} \Text(119,-30)[]{Figure 11: The leading order
diagrams giving a contribution to the $\psi$ field propagator}
\Text(3,-43)[]{and the triple vertex  in $1/N_f$ expansion}
\end{picture}
\end{center}

Calculating the singular parts of the diagrams of Fig.11 in
dimensional regularization with $D'=D-2\varepsilon$ one finds
\begin{eqnarray}\label{sing_ferm}
Diag.a &\Rightarrow& \frac{1}{\varepsilon N_f} \frac{h}{1+h} A,\ \ \ Diag.b\ \Rightarrow\
\frac{1}{\varepsilon N_f} \frac{h}{1+h} B,\ \ \
Diag.c \ \Rightarrow\ \frac{1}{\varepsilon N_f} \frac{h}{1+h} C , \nonumber \\
&& \nonumber \\ &&A=\frac{\Gamma(D)(D-1)(2-D/2)C_F}{2^{[D/2]+1}
\Gamma(2-D/2) \Gamma(D/2+1) \Gamma^2(D/2)T} , \\
&&  B=-\frac{C_F-C_A/2}{C_F}A, \ \ \ C=-\frac{(1-D/2)C_A}{2(2-D/2)C_F}A \nonumber,
\end{eqnarray}
which is again in agreement with~\cite{Gracey2}. Notice that the third diagram is
proportional to  $h/(1+h)$ instead of $h^2/(1+h)^2$ as in the scalar case. The reason is
that now we have an effective triple gauge vertex proportional to $\sqrt{h}(1+h)$ instead
of $\sqrt{h}\ h$ that cancels one factor of $h/(1+h)$.

Therefore, in the leading order in the $1/N_f$ expansion the renormalization constants
take the form
\begin{eqnarray}\label{r3}
Z_2^{-1}&=&1-\frac{1}{\varepsilon N_f}\frac{Ah}{(1+h)},\\
Z_1&=&1-\frac{1}{\varepsilon N_f}\frac{(B+C)h}{(1+h)}, \\
Z_h&=& Z_1^2Z_2^{-2}=1-\frac{1}{\varepsilon N_f }\frac{2(A+B+C)h}{(1+h)}.
\end{eqnarray}

To check the gauge invariance, we calculated the renormalization
of the coupling through the gauge-ghost interaction.  The leading
diagrams are shown in Fig.12.

\begin{center}
\begin{picture}(250,130)(0,-55)
\DashArrowLine(-10,0)(10,0){2} \DashArrowLine(10,0)(40,0){2}
\DashArrowLine(40,0)(60,0){2} \SetWidth{1.5} \GlueArc(25,0)(15,0,180){1}{6}
\SetWidth{0.5} \Text(25,-10)[]{a}

\SetWidth{1.5} \Gluon(110,55)(110,35){1}{3} \SetWidth{0.5}
\DashArrowLine(80,-0)(110,35){2} \DashArrowLine(110,35)(140,-0){2} \SetWidth{1.5}
\Gluon(89,10)(131,10){1}{6} \SetWidth{0.5} \Text(110,-10)[]{b}

\DashArrowLine(170,0)(190,0){2} \DashArrowLine(190,0)(230,0){2}
\DashArrowLine(230,0)(250,0){2} \SetWidth{1.5} \Gluon(190,0)(210,35){1}{6}
\Gluon(230,0)(210,35){1}{6}
\Vertex(210,35){3} \Gluon(210,35)(210,60){1}{3} \SetWidth{0.5}
\Text(210,-10)[]{c} \Text(131,-30)[]{Figure 12: The leading order
diagrams giving a contribution to the ghost field propagator}
\Text(3,-43)[]{and the triple vertex  in $1/N_f$ expansion}
\end{picture}
\end{center}
Calculating the singular parts of the diagrams  in dimensional regularization one finds
\begin{eqnarray}\label{sing_ghost}
Diag.a &\Rightarrow& \frac{1}{\varepsilon N_f} \frac{h}{1+h}A',\ \ Diag.b\ \Rightarrow\
\frac{1}{\varepsilon N_f} \frac{h}{1+h} B',\
\ Diag.c \ \Rightarrow\ \frac{1}{\varepsilon N_f} \frac{h}{1+h} C' ,  \nonumber\\
&& \nonumber \\ &&\hspace*{-2.5cm} A'=\frac{\Gamma(D)(D-1)C_A}{2^{[D/2]+2} \Gamma(2-D/2)
\Gamma(D/2+1) \Gamma^2(D/2)T} , \ \ \ B'=0, \ \ C'=0,
\end{eqnarray}
which gives the following renormalization constants in the ghost sector
\begin{eqnarray}\label{ghost}
\widetilde{Z}_1&=&1,  \\
\widetilde{Z}_2^{-1}&=&1-\frac{1}{\varepsilon N_f}\frac{A'h}{1+h},\\
Z_h&=&\widetilde{Z}_1^2\widetilde{Z}_2^{-2}=1-\frac{2}{\varepsilon N_f}\frac{A'h}{1+h}.
\end{eqnarray}
One can see that the following relation holds:
\begin{equation}\label{unit}
 A+B+C \ = \ A'+B'+C',
\end{equation}
which follows from the gauge invariance.

We look now at the next-to-leading order to compare it with the
scalar case. The corresponding diagrams for the fermion propagator
are shown in Fig.13. They require some explanation. The first line
of diagrams in Fig.13 is obtained from the one-loop diagrams of
Fig.11 by inserting into the vertex or the fermion line of the
one-loop divergent subgraphs from Fig.11. For example, the diagram
$d$ in Fig.13 is the diagram $a$ from Fig.11 with divergent
one-loop subgraph $c$ from Fig.11 substituted instead of the
initial vertex. The second line of the diagrams in Fig.13 is
obtained from the "forbidden" diagram of Fig.14 by inserting the
same one-loop divergent subgraphs from Fig.11. The diagram $e$ is
the diagram of Fig.14 with insertion of the subgraph $a$ from
Fig.11 into the fermion line (see Fig.15) and the diagram $g$
comes from the insertion of the subgraph $c$ from Fig.11 instead
of one of the vertices in the fermion loop (see Fig.16).


\begin{center}
\begin{picture}(400,235)(0,-170)

\ArrowLine(0,0)(10,0) \ArrowLine(10,0)(20,0)\ArrowLine(20,0)(70,0)
\ArrowLine(60,0)(70,0)\ArrowLine(70,0)(80,0) \SetWidth{1.5}
\GlueArc(40,0)(20,0,180){1}{9} \GlueArc(40,0)(30,0,180){1}{11} \SetWidth{0.5}
\Text(40,-30)[]{a}

\ArrowLine(100,0)(110,0) \ArrowLine(110,0)(130,0) \ArrowLine(130,0)(150,0)
\ArrowLine(150,0)(170,0) \ArrowLine(170,0)(180,0) \SetWidth{1.5}
\GlueArc(130,0)(20,0,180){1}{7} \GlueArc(150,0)(20,180,360){1}{7} \SetWidth{0.5}
\Text(140,-30)[]{b}

\ArrowLine(200,0)(210,0) \ArrowLine(210,0)(230,0) \ArrowLine(230,0)(250,0)
\ArrowLine(250,0)(270,0) \ArrowLine(270,0)(280,0) \SetWidth{1.5}
\GlueArc(220,0)(10,0,180){1}{6} \GlueArc(260,0)(10,0,180){1}{6} \SetWidth{0.5}
\Text(240,-30)[]{c}

\ArrowLine(300,0)(320,0) \ArrowLine(320,0)(340,0) \ArrowLine(340,0)(360,0)
\ArrowLine(360,0)(380,0) \SetWidth{1.5} \Gluon(320,0)(340,40){1}{6}
\Gluon(340,0)(340,40){1}{5} \Gluon(360,0)(340,40){1}{6} \Vertex(340,40){3} \SetWidth{0.5}
\Text(340,-30)[]{d}

\ArrowLine(0,-100)(10,-100) \ArrowLine(10,-100)(90,-100) \ArrowLine(90,-100)(100,-100)
\Oval(50,-65)(15,20)(0) \ArrowArc(50,-65)(15,269,271) \ArrowArc(50,-65)(15,89,91)
\SetWidth{1.5} \GlueArc(50,-100)(40,120,180){1}{5} \GlueArc(50,-100)(40,0,60){1}{5}
\GlueArc(50,-45)(15,-143,-37){1}{4} \SetWidth{0.5} \Text(50,-120)[]{e}

\ArrowLine(120,-100)(130,-100) \ArrowLine(130,-100)(210,-100)
\ArrowLine(210,-100)(220,-100) \Oval(170,-65)(15,20)(0) \ArrowArc(170,-65)(16.5,134,136)
\ArrowArc(170,-65)(16.5,44,46) \ArrowArc(170,-65)(16.5,224,226)
\ArrowArc(170,-65)(16.5,314,316) \SetWidth{1.5} \GlueArc(170,-100)(40,120,180){1}{5}
\GlueArc(170,-100)(40,0,60){1}{5} \Gluon(170,-50)(170,-80){1}{5} \SetWidth{0.5}
\Text(170,-120)[]{f}

\ArrowLine(250,-100)(260,-100) \ArrowLine(260,-100)(340,-100)
\ArrowLine(340,-100)(350,-100) \SetWidth{1.5} \GlueArc(280,-100)(20,90,180){1}{5}
\GlueArc(320,-100)(20,0,90){1}{5}
\Gluon(280,-80)(310,-70){1}{5} \Gluon(280,-80)(310,-90){1}{5} \Vertex(280,-80){3}
\SetWidth{0.5} \ArrowLine(320,-80)(310,-70) \ArrowLine(310,-90)(320,-80)
\ArrowLine(310,-70)(310,-90) \Text(300,-120)[]{g}

\Text(196,-140)[]{Figure 13: The second order diagrams giving a
contribution to the fermion propagator} \Text(31,-152)[]{in the
$1/N_f$ expansion}
\end{picture}
\end{center}

\begin{center}
\begin{picture}(110,95)(0,-35)
\ArrowLine(-10,0)(10,0) \ArrowLine(10,0)(90,0)
\ArrowLine(90,0)(110,0) \Oval(50,35)(15,20)(0)
\ArrowArc(50,35)(15,89,91) \ArrowArc(50,35)(15,269,271)
\SetWidth{1.5} \GlueArc(50,0)(40,120,180){1}{7}
\GlueArc(50,0)(40,0,60){1}{7} \SetWidth{0.5}
\Text(50,-20)[]{Figure 14: The "forbidden" loop diagram}
\end{picture}
\end{center}

\begin{center}
\begin{picture}(270,95)(0,-35)
\ArrowLine(-10,0)(10,0) \ArrowLine(10,0)(90,0) \ArrowLine(90,0)(110,0)
\Oval(50,35)(15,20)(0) \ArrowArc(50,35)(15,89,91) \ArrowArc(50,35)(15,269,271)
\SetWidth{1.5} \GlueArc(50,0)(40,120,180){1}{7} \GlueArc(50,0)(40,0,60){1}{7}
\SetWidth{0.5}

\Text(112,20)[]{+}

\ArrowLine(120,10)(160,10)\SetWidth{1.5} \GlueArc(140,10)(15,0,180){1}{7} \SetWidth{0.5}

\Text(165,20)[]{$\rightarrow$}

\ArrowLine(170,0)(180,0) \ArrowLine(180,0)(260,0) \ArrowLine(260,0)(270,0)
\Oval(220,35)(15,20)(0) \ArrowArc(220,35)(15,269,271) \ArrowArc(220,35)(15,89,91)
\SetWidth{1.5} \GlueArc(220,0)(40,120,180){1}{5} \GlueArc(220,0)(40,0,60){1}{5}
\GlueArc(220,55)(15,-143,-37){1}{4} \SetWidth{0.5}

\Text(129,-20)[]{Figure 15: The diagram $e$ from Fig.13 as a
result of insertion of the diagram $a$ from }
\Text(-17,-35)[]{Fig.11 into the fermion line.}
\end{picture}
\end{center}

\begin{center}
\begin{picture}(270,95)(0,-35)
\ArrowLine(-10,0)(10,0) \ArrowLine(10,0)(90,0) \ArrowLine(90,0)(110,0)
\Oval(50,35)(15,20)(0) \ArrowArc(50,35)(15,89,91) \ArrowArc(50,35)(15,269,271)
\SetWidth{1.5} \GlueArc(50,0)(40,120,180){1}{7} \GlueArc(50,0)(40,0,60){1}{7}
\SetWidth{0.5}

\Text(112,20)[]{+}

\ArrowLine(120,0)(160,0) \ArrowLine(130,20)(150,20) \ArrowLine(150,20)(140,30)
\ArrowLine(140,30)(130,20) \SetWidth{1.5} \Gluon(130,0)(130,20){1}{5}
\Gluon(150,0)(150,20){1}{5} \Gluon(140,30)(140,50){1}{5} \SetWidth{0.5}

\Text(165,20)[]{$\rightarrow$}

\ArrowLine(170,0)(180,0) \ArrowLine(180,0)(260,0) \ArrowLine(260,0)(270,0) \SetWidth{1.5}
\GlueArc(200,0)(20,90,180){1}{5} \GlueArc(240,0)(20,0,90){1}{5}
\Gluon(200,20)(230,30){1}{5} \Gluon(200,20)(230,10){1}{5} \Vertex(200,20){3}
\SetWidth{0.5} \ArrowLine(240,20)(230,30) \ArrowLine(230,10)(240,20)
\ArrowLine(230,30)(230,10)

\Text(127,-20)[]{Figure 16: The diagram $g$ from Fig.13 as a
result of insertion of the diagram $c$ from}
\Text(75,-35)[]{Fig.11 instead of one of the vertices in the
diagram from Fig.14.}
\end{picture}
\end{center}
%

All these diagrams are double logarithmically divergent, i.e., contain both single and
double poles in dimensional regularization. We calculated the leading double poles after
subtraction of the divergent subgraphs, i.e., performed  the $R'$-operation. The answer
is:
\begin{eqnarray}\label{sing22}
Diag.a &\Rightarrow& -\frac{1}{\varepsilon^2 N_f^2}\frac{A^2h^2}{2(1+h)^2},\  Diag.b\
\Rightarrow\ -\frac{1}{\varepsilon^2 N_f^2}\frac{ABh^2}{(1+h)^2},\
Diag.c \ \Rightarrow\ -\frac{1}{\varepsilon^2 N_f^2}\frac{A^2h^2}{(1+h)^2}, \nonumber \\
Diag.d &\Rightarrow& -\frac{1}{\varepsilon^2 N_f^2}\frac{ACh^2}{(1+h)^2},\ \ \ Diag.e\
\Rightarrow \frac{1}{\varepsilon^2 N_f^2}\frac{2A^2h^3}{3(1+h)^3},\
\\ Diag.f &\Rightarrow& \frac{1}{\varepsilon^2 N_f^2}\frac{2ABh^3}{3(1+h)^3},
\ \ \ Diag.g  \Rightarrow \frac{1}{\varepsilon^2 N_f^2}\frac{2ACh^3}{3(1+h)^3}.\nonumber
\end{eqnarray}
We performed also the calculation for the fermion-gauge-fermion vertex but do not present
the diagram-by-diagram result because of the lack of space and give only the final
answer.

Having all this in mind we come to the final expressions for the Z factors in the second
order of the $1/N_f$ expansion in the fermion sector:
\begin{eqnarray}\label{newg}
Z_1&=&1-\frac{1}{\varepsilon N_f}\frac{(B+C)h}{1+h} + \frac{1}{\varepsilon^2 N_f^2}\left(
\frac 32\frac{(B+C)^2h^2}{(1+h)^2} + \frac{A(B+C)h^2}{(1+h)^2}\right.\nonumber
\\ &&\left. -\frac {2}{3}\frac{(B+C)^2h^3}{(1+h)^3} - \frac
23\frac{A(B+C)h^3}{(1+h)^3} \right)
+O(\frac{1}{\varepsilon N_f^2}), \\
Z_2^{-1}&=&1-\frac{1}{\varepsilon N_f}\frac{Ah}{1+h}+\frac{1}{\varepsilon^2 N_f^2}\left(
\frac
32\frac{A^2h^2}{(1+h)^2} + \frac{A(B+C)h^2}{(1+h)^2} \right.\nonumber \\
&&\left. - \frac 23\frac{A(A+B+C)h^3}{(1+h)^3} \right)+O(\frac{1}{\varepsilon N_f^2}).
\label{new2g}
\end{eqnarray}
The same calculation in  the ghost sector gives
\begin{eqnarray}\label{newghost}
\widetilde{Z}_1&=&1 ,  \\
\widetilde{Z}_2^{-1}&=&1-\frac{1}{\varepsilon N_f}\frac{A'h}{1+h}+\frac{1}{\varepsilon^2
N_f^2}\left( \frac
32\frac{A'^2h^2}{(1+h)^2}
- \frac 23\frac{A'(A+B+C)h^3}{(1+h)^3} \right)+O(\frac{1}{\varepsilon N_f^2}). \nonumber
\end{eqnarray}
Notice the absence of the ghost-gauge-ghost vertex renormalization.

The final second order expression for the coupling renormalization calculated in both
ways having in mind relation (\ref{unit}) is
\begin{equation}\label{Renorm}
Z_h=1 - \frac {1}{\varepsilon N_f} \frac{2(A+B+C)h}{1+h} + \frac{1}{\varepsilon^2 N_f^2}
  \left(4\frac{(A+B+C)^2h^2}{(1+h)^2} -
\frac43\frac{(A+B+C)^2h^3}{(1+h)^3}\right)+O(\frac{1}{\varepsilon N_f^2}).
\end{equation}

Like in the scalar and QED case, one can also calculate the
renormalization of the original coupling $g^2$. The leading order
diagrams are shown in Fig. 17 which give the following singular
parts like in~\cite{Gracey2}
\begin{center}
\begin{picture}(400,120)(0,-160)

\Oval(50,-65)(15,20)(0) \ArrowArc(50,-65)(15,89,91) \ArrowArc(50,-65)(15,269,271)
\Text(50,-95)[]{a} \SetWidth{1.5} \Gluon(10,-65)(30,-65){1}{3}
\Gluon(70,-65)(90,-65){1}{3} \GlueArc(50,-45)(15,-143,-37){1}{5} \SetWidth{0.5}
\Line(46,-56)(54,-64) \Line(54,-56)(46,-64)

\Oval(160,-65)(15,20)(0) \ArrowArc(160,-65)(16.5,44,46) \ArrowArc(160,-65)(16.5,134,136)
\ArrowArc(160,-65)(16.5,224,226) \ArrowArc(160,-65)(16.5,314,316) \SetWidth{1.5}
\Gluon(120,-65)(140,-65){1}{3} \Gluon(180,-65)(200,-65){1}{3}
\Gluon(160,-50)(160,-80){1}{3} \SetWidth{0.5} \Line(156,-61)(164,-69)
\Line(164,-61)(156,-69) \Text(160,-95)[]{b}

\SetWidth{1.5} \Gluon(220,-65)(240,-65){1}{3} \Gluon(280,-65)(300,-65){1}{3}
\GlueArc(260,-65)(20,0,180){1}{7} \GlueArc(260,-65)(20,180,360){1}{7} \Vertex(240,-65){3}
\Vertex(280,-65){3} \SetWidth{0.5} \Line(256,-39)(264,-49) \Line(264,-39)(256,-49)
\Text(260,-95)[]{c}

\SetWidth{1.5} \Gluon(320,-65)(340,-65){1}{3} \Gluon(380,-65)(400,-65){1}{3}
\GlueArc(360,-65)(20,0,180){1}{7} \GlueArc(360,-65)(20,180,360){1}{7} \Vertex(380,-65){3}
\SetWidth{0.5} \Line(346,-60)(334,-70) \Line(334,-60)(346,-70) \Text(360,-95)[]{d}

\Text(192,-115)[]{Figure 17: The first order diagrams giving a
contribution to the $1/g^2$ renormalization} \Text(22,-130)[]{in
$1/N_f$ expansion}
\end{picture}
\end{center}\vspace{-1cm}

\begin{eqnarray}\label{lambda2}
\nonumber Diag.a &\Rightarrow& \frac{1}{\varepsilon N_f} \frac{h^2}{(1+h)^2}F,\ \ Diag.b\
\Rightarrow\ \frac{1}{\varepsilon N_f} \frac{h^2}{(1+h)^2}E,  \\  Diag.c & \Rightarrow&
\frac{1}{\varepsilon N_f} \frac{h^2}{(1+h)^2}G, \ \ \ Diag.d\ \Rightarrow\
\frac{1}{\varepsilon N_f} \frac{h^2}{(1+h)^2}H,  \\
 \nonumber  F&=&\frac{\Gamma(D+1)(D/2-1)(D-1)^2(2-D/2)C_F}{2^{[D/2]+2}
\Gamma(2-D/2) \Gamma(D/2+2) \Gamma^2(D/2)T(R)}, \\
 E&=&-\frac{D^2+D/2-9}{D/2(D/2-1)(D-1)}\frac{C_F-C_A/2}{C_F}F, \nonumber \\
 \nonumber
G&=&\frac{4(D/2)^6-6(D/2)^5+18(D/2)^4-67(D/2)^3+85(D/2)^2-19D+6}{2(D-1)^2(1-D/2)^2(2-D/2)D}
\frac{C_A}{C_F} F, \\
 \nonumber  H&=&\frac{D^3-D^2/2-2D+1}{D(1-D/2)(2-D/2)(D-1)^2}\frac{C_A}{C_F }F. \nonumber
\end{eqnarray}
The corresponding renormalization constant looks like
\begin{equation}
Z_{1/ g^2} \ = \ 1 - \frac{1}{\varepsilon N_f} \frac{(F+E+G+H)h^2}{(1+h)^2}.
\end{equation}

\section{Renormalization group in 1/N expansion}

Having these expressions for the Z factors one can construct the coupling constant
renormalization and the  corresponding RG functions. One has as usual in the dimensional
regularization
\begin{eqnarray}\label{rg}
  h_B&=&(\mu^2)^\varepsilon hZ_1^2Z_2^{-2}=(\mu^2)^\varepsilon \left(
  h+\sum_{n=1}^{\infty}\frac{a_n(h,N)}{\varepsilon^n}\right),\\
  Z_i&=&1+\sum_{n=1}^{\infty}\frac{c^i_n(h,N)}{\varepsilon^n},\label{zz}
\end{eqnarray}
where the first coefficients $a_n$ and $c^i_n$ can be deduced from the $Z$ factors.

This allows one to get the anomalous dimensions and the beta function defined as
\begin{eqnarray}\label{anom}
  \gamma(h,N)&=&-\mu^2\frac{d}{d\mu^2}\log Z = h\frac{d}{dh}c_1, \\
 \beta(h,N)&=&2h(\gamma_1+\gamma_2)=(h\frac{d}{dh}-1)a_1.
\end{eqnarray}

We first consider the scalar case.  With the help of
eqs.(\ref{newscalar},\ref{new2}) one gets in the leading order of
$1/N$ expansion\footnote{Note that the anomalous dimension of a
field $\gamma_2$, is defined with respect to $Z_2^{-1}$.}
\begin{eqnarray}\label{dim}
  \gamma_2(h,N)&=&-\frac 1N \frac{Ah}{(1+h)^2}, \ \ \
  \gamma_1(h,N)=-\frac 1N \left(\frac{Bh}{(1+h)^2}+\frac{2Ch^2}{(1+h)^3}\right),\\
  \beta(h,N)&=&-\frac 1N \left(\frac{2(A+B)h^2}{(1+h)^2}+\frac{4Ch^3}{(1+h)^3}\right).
\end{eqnarray}

It is instructive to check the so-called pole equations~\cite{Hooft} that express the
coefficients of the higher order poles in $\varepsilon$ of the Z factors via the
coefficients of a simple pole. For $Z_2^{-1}$ one has, according to (\ref{new2}),
\begin{eqnarray}\label{poles}
  c_1(h,N)&=&-\frac 1N \frac{Ah}{1+h}, \\
  c_2(h,N)&=& \frac{1}{N^2}
  \left(
\frac 32\frac{A^2h^2}{(1+h)^2}+\frac{ABh^2}{(1+h)^2}+\frac
23\frac{A^2h^3}{(1+h)^3}\right.\nonumber \\ &&\left. + \frac 23\frac{ABh^3}{(1+h)^3}
+\frac 43\frac{ACh^3}{(1+h)^3}+\frac{ACh^4}{(1+h)^4}\right).\label{poles2}
\end{eqnarray}
At the same time the coefficient $c_2$ can be expressed through $c_1$ via the pole
equations as
\begin{equation}\label{p2}
  h\frac{dc_2}{dh}=\gamma_2c_1+\beta\frac{dc_1}{dh},
\end{equation}
which gives
$$h\frac{dc_2}{dh}=\frac{1}{N^2}\frac{Ah}{(1+h)^2}\frac{Ah}{1+h}
+\frac{1}{N^2}\left(\frac{2(A+B)h^2}{(1+h)^2}+\frac{4Ch^3}{(1+h)^3}\right)\frac{A}{(1+h)^2}.$$
Integrating this equation one gets for $c_2$ the expression
coinciding with (\ref{poles2}) which was obtained by direct
diagram evaluation. Notice that to get this coincidence the
$h$-dependence in the denominator of
eqs.(\ref{newscalar},\ref{new2}) was absolutely crucial.

We have also checked  the pole equations for the renormalized coupling. They look as
follows
\begin{equation}\label{rg2}
(h\frac{d}{dh}-1)a_n=\beta\frac{da_{n-1}}{dh}.
\end{equation}
In the leading order in $h$ when
$$a_1(h,N)\simeq-\frac{2(A+B)h^2}{N}\ \ \mbox{and} \ \  \beta(h,N)\simeq-\frac{2(A+B)h^2}{N}$$
one should have a geometric progression
$$a_n(h,N)=a_1(h,N)^n.$$
We have checked this relation up to three loops and confirmed its validity.

Having expression for the $\beta$ function one may wonder how the coupling is running.
The crucial point here is the sign of the $\beta$ function. One has
\begin{equation}\label{beta}
\beta(h,N)=-\frac 1N \frac{\displaystyle
4\Gamma(D-2)\left(\frac{2h^2}{(1+h)^2}+\frac{D(D-3)h^3}{(1+h)^3}
\right)}{\Gamma(D/2-2)\Gamma(D/2-1)\Gamma(D/2+1)\Gamma(3-D/2)}.
\end{equation}
It can also be rewritten as
\begin{equation}\label{beta2}
\beta(h,N)=-\frac 1N \frac{\displaystyle
2^{D-1}\Gamma(D/2-1/2)(-)^{(D-1)/2}\left(\frac{2h^2}{(1+h)^2}+\frac{D(D-3)h^3}{(1+h)^3}
\right)}{\Gamma(1/2)\pi\Gamma(D/2+1)},
\end{equation}
that clearly indicates that the theory is UV asymptotically free for $D=2k+1,\ k$ - even
and IR free for $k$-odd. Solution of the RG equation looks somewhat complicated, but for
the small coupling in the leading order it simply equals the usual leading log
approximation
\begin{equation}\label{sol}
  h(t,h) \simeq \frac{h}{1-\beta_0h\log(t)}, \ \ \ \beta_0=-\frac 1N \frac{\displaystyle
2^{D}\Gamma(D/2-1/2)(-)^{(D-1)/2}}{\Gamma(1/2)\pi\Gamma(D/2+1)}.
\end{equation}
For example, for $D=5,7$ the beta function equals $\beta_0=-256/15\pi^2N$ and
$2^{12}/105\pi^2N$, respectively.

We now come to the gauge theories. With the help of eqs.(\ref{newg},\ref{new2g}) one gets
in the leading order of the $1/N_f$ expansion
\begin{eqnarray}\label{dim2}
  \gamma_2(h,N_f)&=&-\frac {1}{N_f} \frac{Ah}{(1+h)^2}, \ \ \
  \gamma_1(h,N_f)=-\frac {1}{N_f} \frac{(B+C)h}{(1+h)^2},\\
  \widetilde{\gamma}_2(h,N_f)&=&-\frac {1}{N_f} \frac{A'h}{(1+h)^2}, \ \ \
  \widetilde{\gamma}_1(h,N_f)=O(\frac{1}{N_f^2}),\\
  \beta(h,N_f)&=&-\frac {1}{N_f} \frac{2(A+B+C)h^2}{(1+h)^2}, \label{betaferm}
\end{eqnarray}

The situation is similar to that in scalar theory. Only the value of coefficients are
different. This, however, does not influence the pole equations. They remain to be valid.

Equation (\ref{Renorm}) gives us the sign of the beta function. In the leading order one
has
\begin{equation}\label{beta111}
\frac{dh}{dt}=\beta(h)= -\frac{\Gamma(D)(D-1)C_A}{2^{[D/2]+2}
\Gamma(2-D/2) \Gamma(D/2+1) \Gamma^2(D/2)N_fT}\frac{h^2}{(1+h)^2},
\end{equation}
which means that contrary to the scalar case (\ref{beta2})
$\beta(h)>0$ for $D=5$, $\beta(h)<0$ for $D=7$ and then alternates
with $D$ as in the scalar case.

Solution to eq.(\ref{beta111}) for small $h$ is again reduced to
the usual one. As for the original couplings, there is no simple
solution either except for the QED case, where the coupling $h$ is
not running and solution of the RG equation for $1/e^2$ with fixed
$h$ is
\begin{equation}\label{power}
\frac{1}{e^2}=\frac{1}{e^2_0}\left(\frac{p^2}{p^2_0}\right)^\gamma, \label{run}
\end{equation}
with the anomalous dimension
$$  \gamma=
\frac{\Gamma(D)(D-1)(D/2-2)(D-3)(D+2)(D-6)}{2^{[D/2]+2}\Gamma(D/2+2)\Gamma^2(D/2)\Gamma(2-D/2)N_f}
\frac{h^2}{(1+h)^3}. $$ The sign of $\gamma$ depends on $D$. For $D=5,7$ $\gamma>0$, for
$D=9$ $\gamma<0$ and then alternates with every odd $D$. Eq.(\ref{power}) reminds the
power law behaviour of the initial coupling in extra dimensions within the Kaluza-Klein
approach~\cite{powerbehav} though anomalous dimension $\gamma$ is different.

\section{Analytical properties and unitarity}

Consider now the analytical properties of the propagator and
related problem of unitarity.  The problem is common to scalar and
gauge theories so for simplicity we concentrate on the sigma field
propagator (\ref{propscalar1}). Besides the cut starting from
$4m^2$ it has poles in the complex $p^2$ plane. Hence, knowing the
analytical structure, one can write down the K\"allen-Lehmann
representation~\cite{KL}.

Let us first consider the massless case (\ref{pol})
\begin{equation}
  D(p^2) = \frac{i}{1/\lambda + f(D)(-p^2)^{D/2-2}}.
\end{equation}
Depending on a sign of $f(D)$ there are two possibilities: either one has a pole at real
axis and (possibly) pairs of complex conjugated poles ($f(D)<0$, D=5,9,...) or one has
only pairs of complex conjugated poles ($f(D)>0$, D=7,11,...) and all the rest appears at
the second Riemann sheet. We consider the cases of $D=5$ and $D=7$ as the nearest
options. One has, respectively,
\begin{equation}\label{4}
D_5(p^2) \ = \ - \frac{2 (256 \pi)^2
\lambda^2}{p^2+(256\pi/\lambda)^2} + \frac{1}{\pi} \int_0^\infty
\frac{dm^2}{p^2-m^2} \frac{256 \pi \lambda^2
\sqrt{m^2}}{(256\pi)^2 + \lambda^2 m^2},
\end{equation}
and
\begin{eqnarray}\label{6}
D_7(p^2) \ &=& \ - \frac 23 \ \frac{ (\frac{8192
\pi^2}{\lambda})^{2/3}e^{\pi i/3}}{p^2+(\frac{8192
\pi^2}{\lambda})^{2/3}e^{-2 \pi i/3}} - \frac 23 \ \frac{(
\frac{8192 \pi^2}{\lambda} )^{2/3}e^{-\pi i/3}}{p^2+(\frac{8192
\pi^2}{\lambda} )^{2/3}e^{2 \pi i/3}} \nonumber
\\ && + \frac{1}{\pi} \int_0^{\infty} \frac{dm^2}{p^2-m^2} \frac{8192 \pi^2
\lambda^2 (m^2)^{3/2}}{\lambda^2 (m^2)^3+(8192 \pi^2)^2}.
\end{eqnarray}
Notice that the continuous spectrum has a positive spectral density and corresponds to
production of real pairs of $\phi$ fields (or pairs of fermions in the gauge case). These
states are present in the original spectrum and cause no problem with unitarity. This
analysis was performed at the tree level in~\cite{Arefeva} and can be extended to any
number  of loops. One can show that all the cuts imposed on diagrams when applying
Cutkosky rules~\cite{Cut} in any order of perturbation theory lead to the usual
asymptotic states on mass shell and no new states appear.

The problem comes with the poles. One can see that the pole terms
come with negative sign and, therefore, correspond to the ghost
states~\cite{Kr}. For $D=5$ one has only one pole at the positive
real semiaxis while for $D=7$ one has a pair of complex conjugated
poles, as shown in Fig.18.
\begin{center}
\begin{picture}(240,130)(0,-70)

\SetWidth{0.5} \Line(0,0)(50,0) \Line(50,-50)(50,50) \SetWidth{2}
\Line(50,0)(100,0) \SetWidth{0.5} \Line(130,0)(180,0) \SetWidth{2}
\Line(180,0)(230,0) \SetWidth{0.5} \Line(180,-50)(180,50)
\Vertex(25,0){2} \Vertex(200,-20){2} \Vertex(200,20){2}
\Text(10,40)[]{D=5} \Text(140,40)[]{D=7} \Text(95,-10)[]{Re}
\Text(60,40)[]{Im} \Text(225,-10)[]{Re} \Text(190,40)[]{Im}
\Text(94,-70)[]{Figure 18: The analytical structure of the
auxiliary field propagator}
\end{picture}
\end{center}

The presence of these ghost states is the drawback of a theory.
They signal of instability of the vacuum state. Indeed, as it was
shown in~\cite{Aref2}, the vacuum might be unstable with respect
to appearance of condensates. This will lead to additional
diagrams similar to those in QCD. However, they do not seem to
improve the situation. Thus, one has either to try to get rid of
ghost poles or to make sure that they do not give a contribution
to physical amplitudes.

Let us first see what happens if one takes a nonzero mass of the $\phi$ field. The
polarization operator then is
\begin{equation}
\Pi(p^2)= - \frac{\Gamma(2-D/2)}{2^{D+1} \pi^{D/2}} \int_0^1 dx
(-p^2x(1-x)+m^2)^{D/2-2}.
\end{equation}
For $D=5$ one has
\begin{equation}\label{84}
\Pi_5(p^2) \ =  \frac{1}{32 \pi^2} m
\left[\frac{4\sqrt{a}+(a-4)\ln(\frac{2-\sqrt{a}}{2+\sqrt{a}})}{8\sqrt{a}}
\right],
\end{equation}
where $a=\frac{p^2}{m^2}$. Since the existence of a pole is
governed by the equation
$$\Pi(p^2)=1/\lambda , $$
one has to check whether this equation is satisfied somewhere in
the complex $p^2$ plane. Remind that for the massless case the
pole exists at $p^2=-(256\pi/\lambda)^2$. In Fig.19, we show the
plot of $\Pi_5(p^2)$ for real $p^2$ (left) and the absolute value
in the complex plane (right).\vspace{1cm}

 \includegraphics[width=200pt,height=180pt,bb=0 0 535 459]{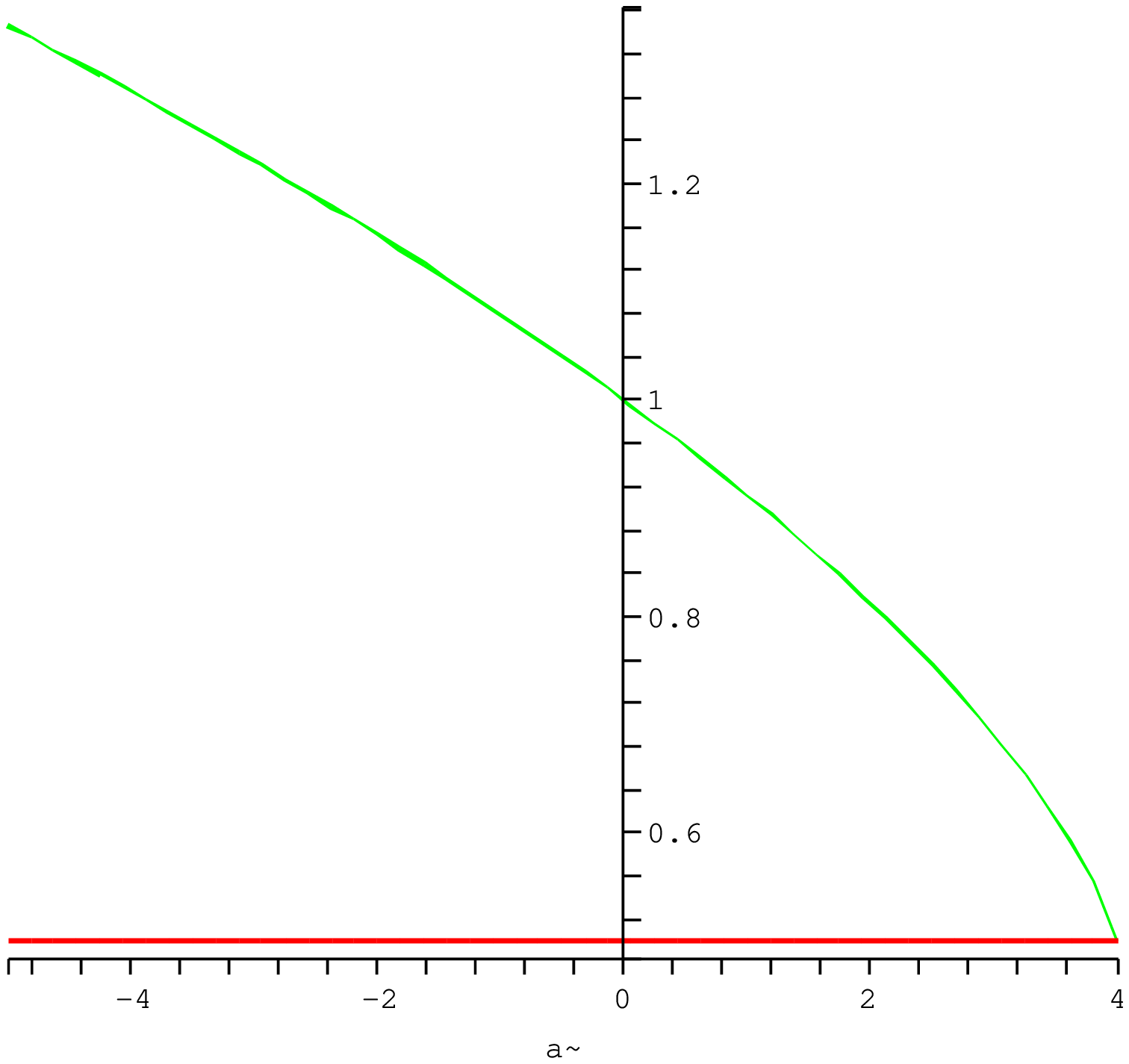}\vspace{-6.6cm}

\hspace*{6.5cm}\includegraphics[width=250pt,height=200pt]{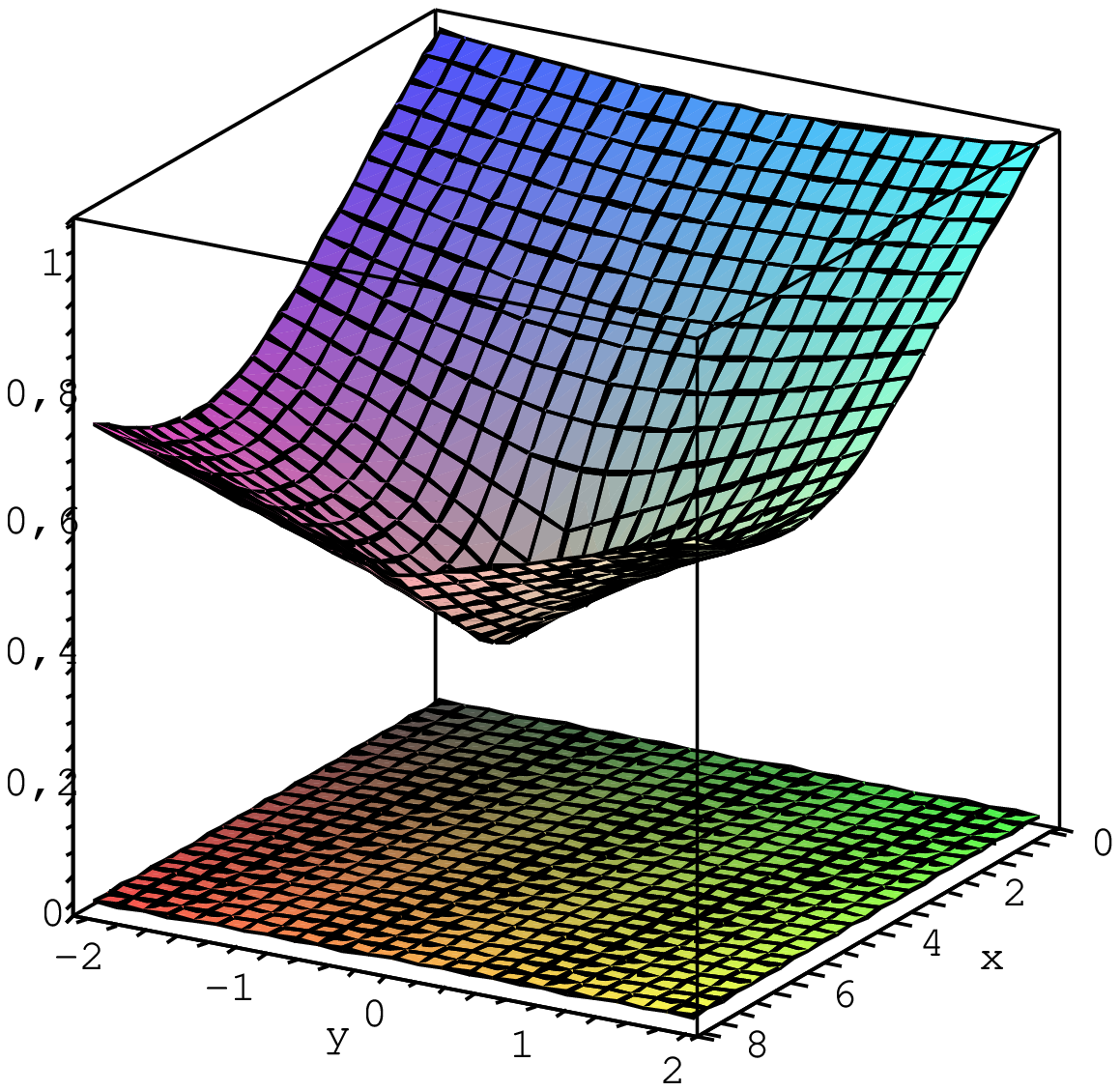}

Fig.19: Polarization operator for D=5 as a function of $p^2/m^2$
for real $p^2$(left) and the absolute value in the complex plane
(right).\vspace{1cm}

One can see that for negative $p^2$ the polarization operator is
always greater than 1 (in units of $\frac{1}{32 \pi^2}m\lambda$),
for positive $p^2$ it is greater than 1/2 and then becomes
complex. The absolute value in the complex plane is also always
greater than 1/2. This means that depending on the value of
dimensionless parameter $\xi=\lambda^2m$ one has different
possibilities: for $\xi<32 \pi^2$ the pole exists at negative
real $p^2$, for 
$32 \pi^2<\xi<64 \pi^2$ the pole exists at positive real
$p^2<4m^2$. For 
$\xi>64 \pi^2$ there are no poles at all. In this phase a theory
is free from unphysical states. A similar situation, but in 4
dimensions, was discussed in~\cite{Schnitzer}.

So, it looks like by choosing parameter $\xi$ one can get rid of the unitarity problem.
However, it reappears the other way. Indeed, one can see that the denominator of the
propagator in this phase becomes negative. It is also negative at $p^2=0$. In the scalar
case the value of the $\sigma$ field propagator at $p^2=0$ defines the effective
potential of $\phi$ fields after integrating out the auxiliary field $\sigma$. This way
the negative value of the propagator leads to effective potential with negative quartic
coupling, i.e. unbounded from below. In the case of the gauge theory the value of the
denominator at $p^2=0$ defines the sign of the residue of the gauge field propagator at
$p^2=0$, i.e. the metric of the gauge field. Negative sign apparently leads to the
"wrong" metric which is also not acceptable. Thus, the presence of a pole at the real
axis is certainly a problem.

The situation is different in $D=7$ dimensions.  Here one has
\begin{equation}\label{85}
\Pi_7(p^2)\ = -\frac{1}{192 \pi^3} m^{3/2}
\left[\frac{4\sqrt{a}(20-3a)-3(a-4)^2
\ln(\frac{2-\sqrt{a}}{2+\sqrt{a}})}{128\sqrt{a}} \right].
\end{equation}
Notice the sign difference compared to the $D=5$ case which  means
that here there is no pole in the Euclidian region but in the
complex plane. In Fig.20, we present the same plots as above but
for D=7. \vspace{1cm}

 \includegraphics[width=190pt,height=180pt]{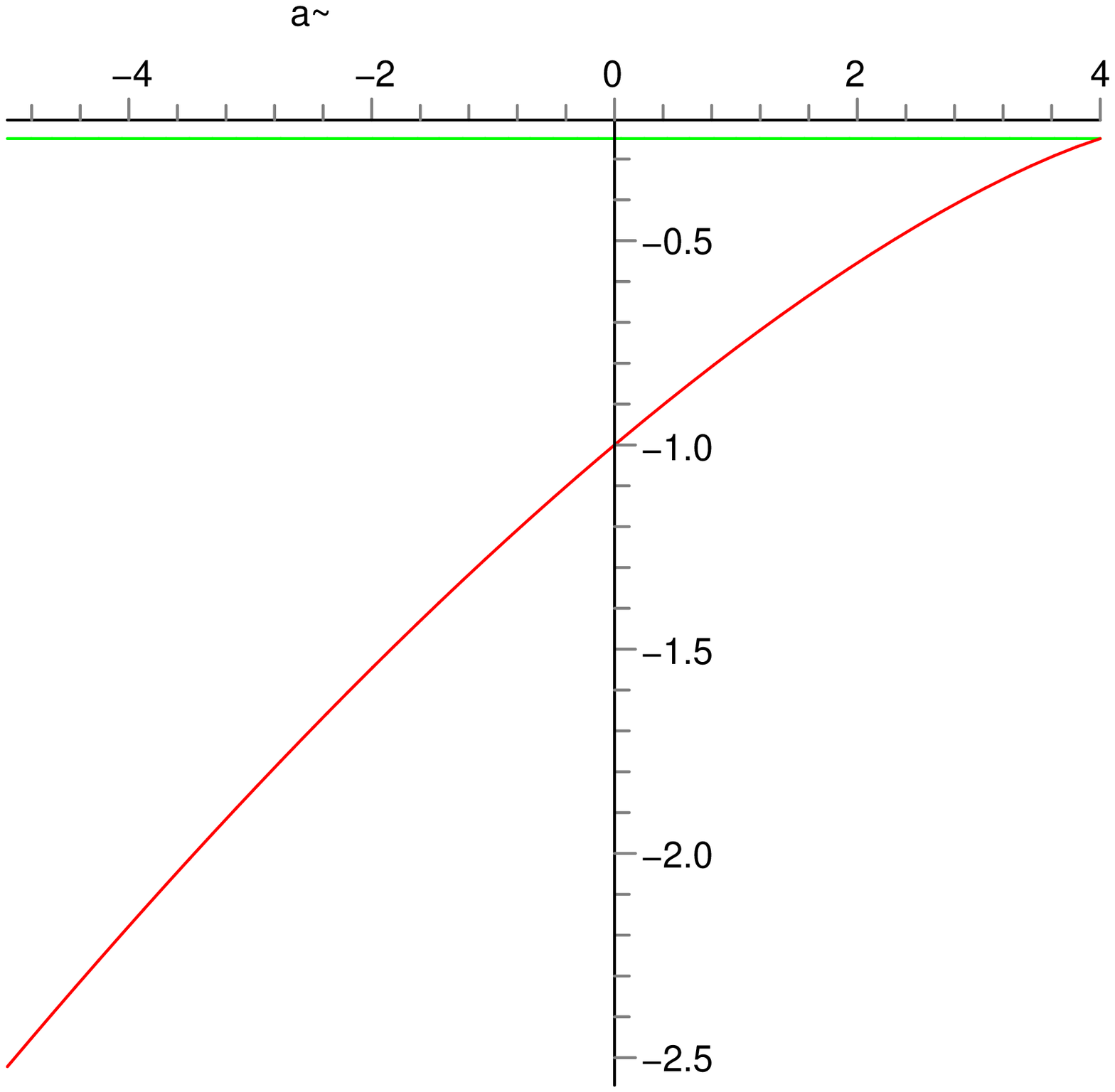}\vspace{-6.8cm}

\hspace*{6.5cm}\includegraphics[width=250pt,height=220pt]{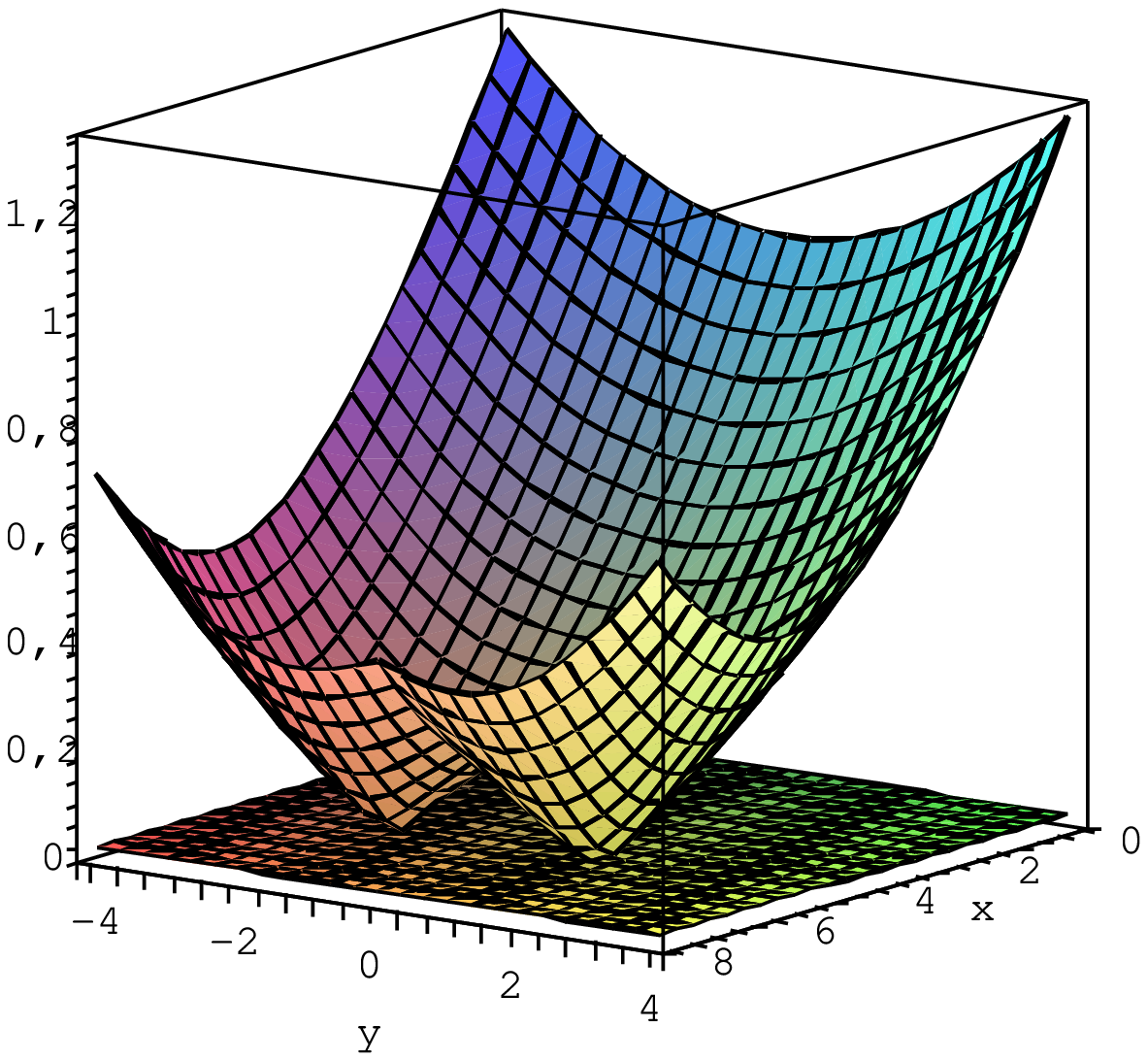}

Fig.20: Polarization operator for D=7 as a function of $p^2/m^2$
for real $p^2$(left) and the absolute value in the complex plane
(right).\vspace{0.6cm}

One can see that the polarization operator in this case can take
any value including negative ones. This means that complex
conjugate poles exist for any value of $\xi=\lambda m^{3/2}$. As
was already mentioned, they correspond to the ghost states and
create trouble unless they are canceled.

We now come to the last step of our analysis. According to
Ref.\cite{Tomboulis}, in the leading order the contribution of
complex conjugated poles to a physical amplitude is canceled, thus
preserving the unitarity in physical subspace. To check this, we
consider the D=7 case and calculate the contribution from the
conjugated ghost poles to the imaginary parts of the Feynman
diagrams in the leading and next-to-leading order of the $1/N$
expansion.

Consider first the one loop diagram shown in Fig.21a. From the
K\"allen-Lehmann representation for the propagator of the
auxiliary field (\ref{6}) we take only the ghost terms ignoring
the continuous spectrum. It corresponds to the following integral:

\begin{picture}(110,50)(-70,0)
\ArrowLine(-10,0)(10,0) \ArrowLine(10,0)(90,0) \ArrowLine(90,0)(110,0) \SetWidth{1.5}
\GlueArc(50,0)(40,0,180){1}{14} \SetWidth{0.5}\Text(50,-20)[]{a)}
\end{picture}\vspace{-1.95cm}

\hspace*{9cm} \begin{picture}(110,95)(0,-40) \ArrowLine(-10,0)(10,0)
\ArrowLine(10,0)(90,0) \ArrowLine(90,0)(110,0) \SetWidth{1.5}
\GlueArc(50,0)(40,0,180){1}{14} \GlueArc(50,0)(20,0,180){1}{7} \SetWidth{0.5}
\Text(50,-20)[]{b)}
\end{picture}\vspace{-0.2cm}

Fig.21: The leading order and next-to-leading order propagator
diagrams\vspace{0.2cm}

\begin{equation}\label{31}
\int \frac{dk}{(k-p)^2} \left(  \frac{R}{k^2-M^2} + \frac{R^*}{k^2-M^{2*}} \right),
\end{equation}
where $R$ and $R^*$ are complex numbers and $M^2$, $M^{2*}$ are the masses of the
conjugated ghost states.

After integration, according to the dimensional regularization
prescription, one gets
\begin{equation}\label{32}
R \int_0^1 dx (1-x)^{3/2}(p^2x+M^2+i\Gamma)^{3/2} + R^* \int_0^1 dx
(1-x)^{3/2}(p^2x+M^2-i\Gamma)^{3/2}.
\end{equation}

For real $p^2$ the integrand apparently has no imaginary part
being the sum of two complex conjugated expressions. The
integration does not change this property: the contribution of the
ghost states to the imaginary part (to physical amplitude) is
canceled and only the continuous spectrum remains. What is crucial
here is that the ghost states are conjugated having the opposite
sign of the imaginary part and the same real part.

Consider now the next-to-leading order diagram shown in Fig.15b. In this diagram there
are several ways how the ghosts might enter
\begin{enumerate}
  \item the inner propagator - non-ghosts, the outer propagator -
  ghosts;
  \item the inner propagator - ghosts, the outer propagator -
  non-ghosts;
  \item the inner propagator - ghosts, the outer propagator -
  ghosts.
\end{enumerate}

Consider the case when the ghost modes run in the inner propagator
and in the outer propagator there is a continuous spectrum
(non-ghosts). Then, using the integral representation (\ref{6})
one has the expression
\begin{eqnarray}\label{34}
&& \sim  \int dm^2 \frac{(m^2)^{3/2}}{\lambda^2(m^2)^3 +
(8192\pi^2)^2} \\
&& \int \frac{dk dq}{(k^2-m^2)((p-k)^2)^2(p-k-q)^2} \left(
\frac{R}{q^2-M^2} + \frac{R^*}{q^2-M^{2*}} \right). \nonumber
\end{eqnarray}

Let us first take the integral  over $q$. One has for the ghost
part (and similar for the conjugated one)
\begin{equation}\label{35}
\sim R \int \frac{dk}{(k^2-m^2)((p-k)^2)^2} \int_0^1 dx
((p-k)^2x(1-x)+M^2x)^{3/2-\varepsilon},
\end{equation}
where we keep $\varepsilon$ to be finite since the two-loop
integral diverges and omit the spectral integration over $m^2$.
The latter is real and is inessential.

Evaluating the integral over $k$ one gets
\begin{eqnarray}\label{37}
&& R \int_0^1 (x(1-x))^{-1/2-\varepsilon} (1-x)^2 dx \int_0^1 dy \int_0^y dz (y-z)
z^{-1/2+\varepsilon}  \nonumber \\  \{  &+&
\Gamma(2\varepsilon)(p^2y(1-y)+m^2(1-y)+M^2\frac{1-x}{x}z+i\Gamma\frac{1-x}{x}z)^{-2\varepsilon}
 \nonumber\\
&&
 \left[
(p^2)^2(1-y)^4x^2+2p^2(M^2+i\Gamma )(1-y)^2x (M^2+i\Gamma )^2\right] \nonumber \\
 &+&  \Gamma(-1+2\varepsilon)
 (p^2y(1-y)+m^2(1-y)+M^2\frac{1-x}{x}z+i\Gamma\frac{1-x}{x}z)^{1-2\varepsilon}
\nonumber\\
 &&  \left[ (7-2\varepsilon+2)p^2x^2(1-y)^2 + (7-2\varepsilon)(M^2+i\Gamma )x \right]
 \nonumber \\
&+& \Gamma(-2+2\varepsilon)
(p^2y(1-y)+m^2(1-y)+M^2\frac{1-x}{x}z+i\Gamma\frac{1-x}{x}z)^{2-2\varepsilon}
\nonumber \\
&&  x^2 \frac{(7-2\varepsilon)(3-2\varepsilon)}{4} \}. \nonumber
\end{eqnarray}
Expanding over $\varepsilon$ one has singular and regular parts. The singular part is
\begin{eqnarray}\label{38}
&& \frac{R}{2\varepsilon} \int_0^1 (x(1-x))^{-1/2} (1-x)^2 dx \int_0^1 dy \int_0^y dz
(y-z) z^{-1/2}  \nonumber \\  \{ &+&
 \left[
(p^2)^2(1-y)^4x^2+2p^2(M^2+i\Gamma )(1-y)^2x (M^2+i\Gamma )^2\right] \nonumber \\
 &-&
 (p^2y(1-y)+m^2(1-y)+M^2\frac{1-x}{x}z+i\Gamma\frac{1-x}{x}z)
 \left[ 9p^2x^2(1-y)^2 + 7(M^2+i\Gamma )x \right]
 \nonumber \\
&+& (p^2y(1-y)+m^2(1-y)+M^2\frac{1-x}{x}z+i\Gamma\frac{1-x}{x}z)^{2}x^2 \frac{21}{8} \}.
\nonumber
\end{eqnarray}
The remaining integrals over Feynman parameters are convergent and can be easily
evaluated. One can see that due to the presence of $i\Gamma$ the integrand is complex,
but adding the complex conjugated term  one gets the real polynomial of $p^2$.

As for the regular part, it contains
$$\log((p^2y(1-y)+m^2(1-y)+M^2\frac{1-x}{x}z+i\Gamma\frac{1-x}{x}z))$$
and has a cut in momentum plane. However, the logarithm can always
be presented in the form $\log(Ae^{i\phi})=\log(A)+i\phi$, where
both the modulus $A$ and the phase $\phi$ depend on Feynman
parameters. This means that adding the conjugated part one again
gets the real integrand and, hence, the real function after
integration. Here it is again crucial that the ghost states are
conjugated and differ only by the sign of the imaginary part.

Thus, we conclude that the contribution from the conjugated ghost
states to the imaginary part of the diagram  is canceled and,
therefore, the ghost states do not contribute to physical
amplitudes. The same analysis can be carried out for the other
choices of the ghost fields in Fig.21b. Moreover, it seems to work
in any diagram in all orders of the $1/N$ expansion since the
reason for the cancellation is simple and obvious. This means that
the unitarity in the physical sector is preserved.

The situation is somewhat similar to that in Ref.~\cite{Slavnov},
where the mass generation problem was discussed in the context of
higher derivative theory.  Besides the physical states there exist
non-physical states with a negative norm, but in the asymptotic
states the negative norm excitations disappear thus preserving the
unitarity of the theory.

\section{Conclusion}

We conclude that in higher dimensional scalar and gauge theories despite  formal
non-renormalizability it is possible to construct renormalizable $1/N_f$ expansion which
obeys all the rules of a usual perturbation theory. The expansion parameter is
dimensionless, the coupling is running logarithmically, all divergencies are absorbed
into the renormalization of the wave function and the coupling.  The original
dimensionful coupling plays a role of a mass and is renormalized multiplicatively.
Expansion over this coupling is singular and creates the usual nonrenormalizable terms.

Properties of the $1/N_f$ expansion do not depend on the space-time dimension if it is
odd. In even dimensions our formulas after subtraction contain a logarithm which creates
some technical problems in calculations but principally do not differ from the odd
dimensions.

Since the actual expansion parameter is dimensionless, all the
Green functions get logarithmic radiative corrections and the
cross-sections decrease with energy like in usual renormalizable
theories without violating the unitarity limit. The running of the
couplings depends on dimension and does not depend on Abelian or
non-Abelian nature of a theory. This may be considered as a
drawback of the $1/N_f$ expansion. Unfortunately, the preferable
$1/N_c$ expansion cannot be constructed in the same simple manner.

We have demonstrated how one can deal with the problem of unitarity and unphysical pole
states. The poles at the real axis can be removed by a proper choice of a dimensionless
parameter $\xi=\lambda m^{D/2-2}$ which corresponds to the correct choice of the phase of
a theory. However, this does not make a theory reliable. At the same time, the complex
conjugated poles remain but fortunately their contribution to the physical amplitudes is
canceled. We do not provide a rigorous proof of this cancellation but present the  reason
for it and several examples how it works in Feynman diagrams. Accepting this reasoning
the theory seems to be unitary in physical subspace.

We hope that this approach can be used in extra dimensional theories to get the
scattering amplitudes. We expect that the behaviour of the cross-sections will differ
from those of the Kaluza-Klein approach~\cite{review} being closer to our
approach~\cite{KV} based on the fixed points.

Besides the already mentioned papers~\cite{Tomboulis} there are
several attempts to build renormalizable effective quantum gravity
using a kind of $1/N$ expansion~\cite{gravity}, where the role of
an expansion parameter $1/N$ is played by the number of space-time
dimensions. The large D limit in this case is very similar to the
large $N_c$ planar diagram limit in the Yang-Mills theory
considered by 't Hooft~\cite{Hooft1}. The technique similar to the
$1/N_f$ expansion is used also in~\cite{Ward}, where the author
sums up the soft graviton corrections to the propagator of the
scalar field and gets an improved propagator which decreases
faster than any power of momenta. Though this partial resummation
is similar to the $1/N$ expansion, the absence of an expansion
parameter does not justify, to our mind, the selected set of
diagrams. From this point of view the $1/N$ expansion is more
consistent and contains the guiding line for such a selection.

\section*{Acknowledgements}

Financial support from RFBR grant \# 05-02-17603 and grant of the Ministry of Education
and Science of the Russian Federation \# 5362.2006.2 is kindly acknowledged. We are
grateful to I. Aref'eva, G. Efimov, A. Kotikov, N. Krasnikov, S. Mikhailov, A. Sheplyakov
and E. Tomboulis for valuable discussions.


\begin{thebibliography}{99}
\bibitem{Original}
N.~Arkani-Hamed, S.~Dimopoulos, and G.R.~Dvali, The Hierarchy problem and new dimensions
at a millimeter, Phys.Lett. {\bf B429}
(1998) 263 [hep-ph/9803315].\\
I.Antoniadis, N. Arkani-Hamed, S. Dimopoulos, and G.R.Dvali, New dimensions at a
millimeter to a Fermi and superstrings at a TeV,
Phys.Lett. {\bf B436} (1998) 257 [hep-ph/9804398];\\
L.Randall and R.Sundrum, An Alternative to compactification, Phys.Rev.Let. {\bf 83}
(1999) 4690 [hep-th/9906064]; A Large mass hierarchy from a small extra dimension,
Phys.Rev.Let. {\bf 83} (1999) 3370 [hep-ph/9905221].
\bibitem{review}
 Yu.A.Kubyshin, Models with Extra Dimensions and Their Phenomenology,
hep-ph/0111027;\\
J.Hewett and M.Spiropulu, Particle physics probes of extra space-time dimensions, Ann.Rev.Nucl.Part.Sci. {\bf 52} (2002) 397 [hep-ph/0205106];\\
D.I.Kazakov, Beyond the standard model, CERN-2006-003 [hep-ph/0411064].
\bibitem{inter} C. Csaki, C. Grojean, H. Murayama, L. Pilo, and J. Terning, Gauge theories on an interval: Unitarity without a Higgs, Phys. Rev. D {\bf 69} 055006 (2004)
[hep-ph/0305237].
\bibitem{our}D.I.Kazakov and G.S.Vartanov, Renormalization group treatment of nonrenormalizable interactions, J.Phys.A:Math.Gen. {\bf 39} (2006) 8051
 [hep-th/0509208].
\bibitem{Kazakov}D.I.Kazakov, On A Generalization Of Renormalization Group Equations To Quantum Field Theories Of An Arbitrary Type, Theor.Math.Phys. {\bf 75} (1988) 440.
\bibitem{KK} G.F.Giudice, R.Rattazzi, J.D.Wells, Quantum gravity and extra dimensions at high-energy colliders, Nucl.Phys. {\bf B544} (1999)
3 [hep-ph/9811291].
\bibitem{esp} E.Alvarez, A.F.Faedo, Renormalized Kaluza-Klein theories, JHEP {\bf 0605} (2006) 046
[hep-th/0602150]; Renormalized masses of heavy Kaluza-Klein states, Phys.\ Rev.\ D {\bf
74} (2006) 124029 [hep-th/0606267]; Quantum corrections to higher-dimensional theories,
hep-ph/0610424.
\bibitem{ratt} R. Rattazzi, Cargese lectures on extra-dimensions, published in Cargese 2003,
Particle physics and cosmology, the Interface, edited by D. Kazakov and G. Smadja, 2005
Springer 461-517 [hep-ph/0607055].
\bibitem{Parisi} G.Parisi, The Theory of Nonrenormalizable Interactions. 1. The Large N Expansion, Nucl.Phys. {\bf B100} (1975) 368.
\bibitem{1N} M.Moshe, J.Zinn-Justin, Quantum field theory in the large N limit: A Review, Phys.Rept. {\bf 385} (2003)
69-228 [hep-th/0306133]; \\
J.Zinn-Justin, Vector models in the large N limit: A Few applications,
SACLAY-SPH-T-97-018 [hep-th/9810198].
\bibitem{Arefeva}I.Ya.Aref'eva, Strong Coupling Limit For The O(N) Phi**4 Interaction, Theor.Math.Phys. {\bf 29} (1976) 147; On The Removal Of Divergences In The Model Of Three-Dimensional N Field, {\it ibid}
 {\bf 31} (1977) 3.
\bibitem{Tomboulis} E. Tomboulis, 1/N Expansion and Renormalization in Quantum Gravity, Phys.Lett. {\bf B70} (1977)
361; Renormalizability And Asymptotic Freedom In Quantum Gravity, Phys.Lett. {\bf B97} (1980) 77; Unitarity In Higher Derivative Quantum Gravity, Phys.Rev.Lett. {\bf 52} (1984) 1173;\\
I.Antoniadis, E. Tomboulis, Gauge Invariance And Unitarity In Higher Derivative Quantum
Gravity, Phys.Rev.D {\bf 33} (1986) 2756.
\bibitem{KVI} D.I.Kazakov and G.S.Vartanov,
Renormalizable Expansion for Nonrenormalizable Theories. I. Scalar Higher Dimensional
Theories, hep-th/0607177.
\bibitem{KVII} D.I.Kazakov and G.S.Vartanov,
Renormalizable Expansion for Nonrenormalizable Theories. II. Gauge Higher Dimensional
 Theories, hep-th/0702004.
\bibitem{reg} G.t'Hooft and M.J.G. Veltman, Regularization and Renormalization of Gauge Fields, Nucl. Phys. {\bf B44} (1972) 189.
\bibitem{Kr} N.V.Krasnikov, Ultraviolet fixed point behavior of the five-dimensional Yang-Mills theory, the gauge hierarchy problem and a possible new dimension at the TeV scale,
Phys.Lett. {\bf B273} (1991) 246; Ultraviolet behavior in five-dimensional Yang-Mills
theory, JETP Lett. {\bf 51} (1990)
4.\\
N.V.Krasnikov and A.B.Kyatkin, Three-dimensional four fermion (vector) x (vector) model
is renormalizable and ultraviolet finite, Mod.Phys.Let.A {\bf 6} (1991) 1315;\\
N.V.Krasnikov, The Four fermion theory is renormalizable, Yad.Phys. {\bf 52} (1990) 1516.
\bibitem{Schnitzer} J.Schnitzer, The 1/N Expansion Of Renormalizable And Nonrenormalizable Scalar Field Theories, Nucl. Phys. {\bf B109} (1976) 297;
\bibitem{Nesterenko} V.V.Nesterenko, On the instability of classical dynamics in
theories with higher derivatives, Phys.Rev. {\bf D75} (2007) 087703 [hep-th/0612265].
\bibitem{Smilga} A.V.Smilga, Ghost-free higher-derivative theory, Phys.Lett. {\bf B632} (2006) 433
[hep-th/0503213]; Benign versus malicious ghosts in higher-derivative theories,
Nucl.Phys. {\bf B706} (2005) 598 [hep-th/0407231].
\bibitem{Antoniadis} I.Antoniadis, E.Dudas, D.M.Ghilencea, Living with ghosts and their radiative corrections, Nucl.Phys. {\bf B767} (2007)
  29 [hep-th/0608094].
\bibitem{Hooft}G.t'Hooft, Dimensional regularization and the renormalization group, Nucl. Phys. {\bf B61} (1973) 455.
\bibitem{renormalon} M.Beneke, Renormalons, Phys.Rep. {\bf 317} (1999) 1 [hep-ph/9807443];\\
M.Beneke and V.Braun, Renormalons and power corrections in B.Ioffe Festschift, World
Scientific, Singapore, 2001, v.3, p. 1719.
\bibitem{Anselmi} D.Anselmi, Large N expansion, conformal field theory and renormalization group flows in three-dimensions, JHEP {\bf 0006} (2000) 042 [hep-th/0005261].
\bibitem{Gracey1} J.A.Gracey, Algorithm for computing the beta function of quantum electrodynamics in the large N(f) expansion, Int. J. Mod. Phys. {\bf A8} (1993)
2465 [hep-th/9301123]; Electron mass anomalous dimension at O(1/(Nf(2)) in quantum
electrodynamics, Phys. Lett. {\bf B317} (1993) 415 [hep-th/9309092].
\bibitem{Gracey2} J.A.Gracey, Quark, gluon and ghost anomalous dimensions at O(1/N(f)) in quantum chromodynamics, Phys. Lett. {\bf B318} (1993) 177
 [hep-th/9310063]; The QCD Beta function at O(1/N(f)), Phys. Lett., {\bf B373} (1996) 178 [hep-ph/9602214].
\bibitem{powerbehav} K.Dienes, E.Dudas, and
T.Gherghetta, Extra space-time dimensions and unification, Phys.Lett. {\bf B436} (1998)
55 [hep-ph/9803466]; Grand unification at intermediate mass scales through extra
dimensions, Nucl.Phys. {\bf B537} (1998) 47 [hep-ph/9806292].
\bibitem{Hooft1} G.t'Hooft, A Planar Diagram Theory for Strong Interactions, Nucl. Phys. {\bf B72} (1974) 461.
\bibitem{KL} G.K\"allen, Helv.Phys.Acta, {\bf 25} (1952) 417,
Quantum Electrodynamics (Springer-Verlag, Berlin, 1972);\\
H.Lehmann, Nuovo Cimento, {\bf 11} (1954) 342.
\bibitem{Cut} R.E.Cutkosky, Singularities and discontinuities of Feynman amplitudes, J.Math.Phys. {\bf 1} (1960) 429.
\bibitem{Aref2}I.Ya.Aref'eva, Phase Transition In The Three-Dimensional Chiral Field, Ann. of Phys. {\bf 117} (1979) 393.
\bibitem{Slavnov} A.A.Slavnov, Higgs mechanism as a collective effect due to extra dimension, Theor.Math.Phys. {\bf 148} (2006) 1159 [hep-th/0604052]; Phys.Lett., {\bf B620} (2005) 97, [arXiv:hep-th/0505195];
  Renormalizable electroweak model without fundamental scalar mesons, hep-th/0601125.
\bibitem{KV} D.I.Kazakov, G.S.Vartanov, On high energy scattering in extra dimensions, Theor.Math.Phys., {\bf 147} (2006) 533 [hep-ph/0410342].
\bibitem{gravity} F. Canfora, The UV behavior of Gravity at Large N, Phys.Rev. {\bf D74} (2006) 064020 [hep-th/0608203]; A Large N expansion for gravity, Nucl.Phys. {\bf
B731} (2005) 389 [hep-th/0511017]; \\
N.E.Bjerrum-Bohr, Quantum gravity at a large number of dimensions,
Nucl.Phys. {\bf B684} (2004) 209 [hep-th/0310263]; \\
A.Strominger, The Inverse Dimensional Expansion In Quantum Gravity, Phys.Rev. {\bf D24}
(1981) 3082;
\bibitem{Ward} B.F.L.Ward, Exact quantum loop results in the theory of general relativity, hep-ph/0607198; Quantum corrections to Newton's law, Mod.Phys.Lett. {\bf A17} (2002) 2371
[hep-ph/0204102]; Are massive elementary particles black holes, ibid. {\bf A19} (2004)
143 [hep-ph/0305058]; J.Cos.Astropart.Phys., {\bf 0402} (2004) 011.
\end{thebibliography}
\end{document}